\documentclass[prd,aps,
nofootinbib,
floatfix,
supserscriptaddress]{revtex4-1}
\usepackage{amssymb}
\usepackage{amsmath,euscript,array,mathrsfs}
\usepackage{graphicx}
\usepackage{dcolumn}
\usepackage{multirow}
\usepackage{bm}
\usepackage{epsfig}
\usepackage{dsfont}
\usepackage{psfrag}
\usepackage{bbold}
\usepackage{rotating}
\usepackage{xcolor}
\usepackage{slashed}
\usepackage[caption=false]{subfig}
\usepackage[all]{xy}
\usepackage[bookmarks=false]{hyperref}
\usepackage{tikz-cd}
\xyoption{import,color,arc}

\newcommand\beq{\begin{eqnarray}}
\newcommand\eeq{\end{eqnarray}}

\newcommand\Eq[1]{Eq.~\ref{eq:#1}}
\newcommand\Fig[1]{Fig.~\ref{fig:#1}}
\newcommand\Sec[1]{Sec.~\ref{sec:#1}}

\newcommand\Tab[1]{Table~\ref{tab:#1}}

\newcommand\gpsi{\gamma_{\bar{\psi}\psi}}
\newcommand\gir{\gamma_{\rm IR}}
\newcommand\girpsi{\gamma_{\bar{\psi}\psi,\,{\rm IR}}}
\newcommand\girpq{\gamma_{{\rm IR},\,[p/q]}}

\newcommand\Gir{\gamma_{\rm IR}(2-\gamma_{\rm IR})}
\newcommand\Girpsi{\gamma_{\bar{\psi}\psi,\,{\rm IR}}(2-\gamma_{\bar{\psi}\psi,\,{\rm IR}})}
\makeatletter

\newcommand{\Rmnum}[1]{\expandafter\@slowromancap\romannumeral #1@}
\makeatother

\begin{document}

\preprint{PNUPT-20/A04}

\title{Conformal window from conformal expansion
}


\begin{abstract}
We study the conformal window of asymptotically free gauge theories containing 
$N_f$ flavors of fermion matter transforming to the vector and two-index representations of $SO(N),~SU(N)$ and $Sp(2N)$ gauge groups. 
For $SO(N)$ we also consider the spinorial representation. 
We determine the critical number of flavors $N_f^{\rm cr}$, 
corresponding to the lower end of the conformal window, 
by using the conjectured critical condition on the anomalous dimension of the fermion bilinear at an infrared fixed point, 
$\girpsi=1$ or equivalently $\Girpsi=1$. 
To compute $\girpsi$ we employ the (scheme-independent) Banks-Zaks conformal expansion up to the $4$th order in $\Delta_{N_f}=N_f^{\rm AF}-N_f$ 
with $N_f^{\rm AF}$ corresponding to the onset of the loss of asymptotic freedom, 
where we show that the latter critical condition provides a better performance along with this conformal expansion.  

To quantify the uncertainties in our analysis, 
which potentially originate from nonperturbative effects, 
we propose two distinct approaches 
by assuming the large order behavior of the conformal expansion separately, 
either convergent or divergent asymptotic. 
In the former case, 
we take the difference in the Pad\'e approximants to the two definitions of the critical condition, 
whereas in the latter case
the truncation error associated with the singularity in the Borel plane 
is taken into account. 
Our results are further compared to other analytical methods as well as lattice results available in the literature. 
In particular, we find that $SU(2)$ with six and $SU(3)$ with ten fundamental flavors 
are likely on the lower edge of the conformal window, 
which are consistent with the recent lattice results. 
We also predict that $Sp(4)$ theories with fundamental and antisymmetric fermions have the critical numbers of flavors, 
approximately ten and five, respectively. 
\end{abstract}

\author{Jong-Wan Lee}
\email{jwlee823@pusan.ac.kr}
\affiliation{Department of Physics, Pusan National University, Busan 46241, Republic of Korea}





\date{\today}

\maketitle
\flushbottom

\section{Introduction}
\label{sec:introduction}

Since it was discovered that $SU(N)$ gauge theories with $N_f$ flavors of fundamental fermions 
could exhibit an interacting conformal phase at an infrared (IR) fixed point 
with a nonzero coupling constant \cite{Caswell:1974gg,Banks:1981nn}, 
a substantial amount of work has been devoted to investigate its properties 
as well as near-conformal behavior in the vicinity of the phase boundary. 
Besides its own theoretical interests, there has also been considerable interest 
in its applications to phenomenological model building for physics beyond the standard model. 
(For instance, see the recent review paper in Ref.~\cite{Cacciapaglia:2020kgq}.) 
In order to access the whole range of the IR conformal phase ({\it conformal window}), 
we typically assume that the number of flavors $N_f$ varies continuously. 
One end of the conformal window is identical to the critical point at which 
the theory loses asymptotic freedom. 
The corresponding number of flavors $N_f^{\rm AF}$ can exactly be determined 
by investigating the renormalization group (RG) beta function in the usual perturbative expansion with respect to the coupling constant.  
At the same time we know that there should be the other end of the conformal window at $N_f^{\rm cr}$ with $0<N_f^{\rm cr}<N_f^{\rm AF}$, 
because the theories with a sufficiently small number of flavors, 
including $N_f=0$ pure Yang-Mills, 
are in the confining phase, and the dynamically generated confinement scale breaks the conformal symmetry. 
In contrast to the upper bound of the conformal window, 
however, it is a highly nontrivial task to identify its lower end 
since we may be in the large coupling regime in general and have to deal with nonperturbative effects. 

Recently, our understanding of the phase structure of non-ablelian gauge theories with fermionic matter 
has been further extended to asymptotically unfree theories for $N_f > N_f^{\rm AF}$ 
\cite{Litim:2014uca,Antipin:2017ebo}.
Just above $N_f^{\rm AF}$ the perturbative beta function yields that the theory possesses a Landau pole, 
and thus it is not well defined in the ultraviolet (UV) while it is trivial in the IR. 
If $N_f$ further increases and becomes larger than $N_f^{\rm safe}$, however, 
the theory develops an ultraviolet fixed point with a nonzero value of the coupling, 
which has been discussed in the context of {\it asymptotic safety} \cite{Litim:2014uca}. 

\begin{figure}
\begin{center}
\includegraphics[width=0.85\textwidth]{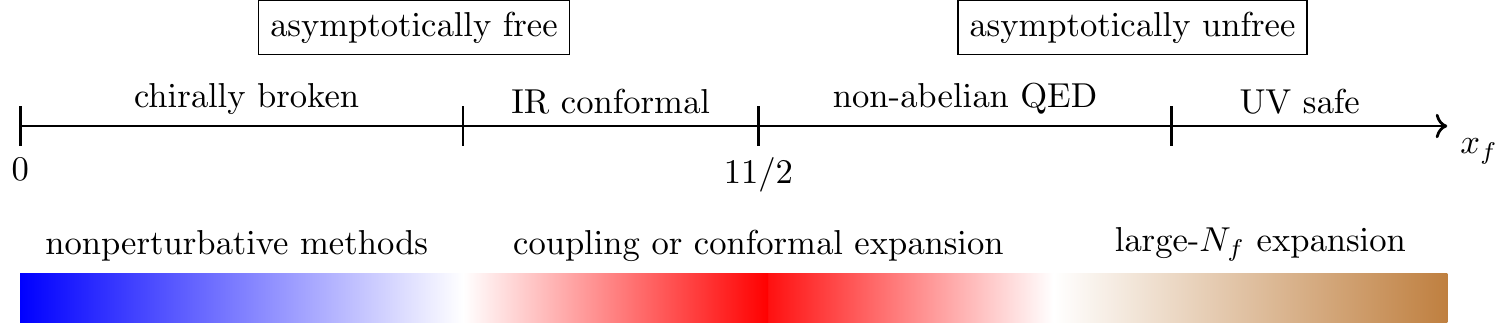}
\caption{
Conjectured phase structure of large $N$ QCD in the Veneziano limit at zero temperature and chemical potential. 
The continuous variable $x_f$ is defined as $x_f=N_f/N$ with both $N_f$ and $N$ taken to be infinite. 
}
\label{fig:large_n_qcd_phase}
\end{center}
\end{figure}

The conjectured phase diagram of non-abelian gauge theories with fermionic matter fields 
at zero temperature and chemical potential 
can then be drawn as in \Fig{large_n_qcd_phase}. 
For illustration purposes we consider that fermions are in the fundamental representation 
and take the large $N$ limit while keeping the ratio $x_f=N_f/N$ is fixed, i.e. the Veneziano limit. 
However, we note that without losing generosity the discussion below can be applied to all the theories 
with a gauge group $G$ and $N_f$ fermions in the representation $R$ considered in this work. 
There are two different phases in which the theory is asymptotically free, chirally broken and IR conformal. 
In the asymptotically unfree regime, two other phases are expected to exist, QED-like and UV safe. 
Analytical understanding of the chirally broken phase at small $x_f$ is highly limited 
because the standard perturbation technique is not applicable 
due to the absence of a small expansion parameter. 
One should instead rely on fully nonperturbative methods such as the lattice Monte-Carlo calculations.

In the vicinity of $x_f^{\rm AF}=11/2$, onset of the loss of asymptotic freedom, 
the coupling expansion of the beta function finds an IR fixed point in the weak coupling regime 
for $x_f < x_f^{\rm AF}$, i.e., IR conformal, but 
it does not for $x_f > x_f^{\rm AF}$ except the Gaussian fixed point at the origin, i.e., non-abelian QED in the IR. 
In this perturbative regime 
one may also consider an alternative series expansion by 
taking the difference, $\Delta_{x_f}\equiv x_f^{\rm AF}-x_f$, as a small parameter. 
Such an expansion, so-called the Banks-Zaks {\it conformal expansion}, has been shown to be a useful tool 
for the investigation of the IR conformal phase \cite{Banks:1981nn}. 
In particular, the scheme-independent conformal expansions of physical quantities at an infrared fixed point, such as the anomalous dimension 
of a fermion bilinear operator $\girpsi$ and the derivative of the beta function $\beta'_{\rm IR}$, 
have been extensively studied in a series of papers 
\cite{Ryttov:2016hdp,Ryttov:2016asb,Ryttov:2016hal,Ryttov:2017toz,Ryttov:2017kmx,Ryttov:2017dhd,Ryttov:2017lkz}.\footnote{%
More work on the conformal expansions of the anomalous dimensions of baryon operators and higher-spin operators, 
and of $\girpsi$ and $\beta'_{\rm IR}$ in the theories with multiple fermion representations can be found in 
Refs.~\cite{Gracey:2018oym,Ryttov:2020scx,Ryttov:2018uue}. 
See also Refs.~\cite{Grunberg:1992mp,Caveny:1997yr,Gardi:1998ch} for some earlier work on the conformal expansion 
of $\beta'_{\rm IR}$ in QCD. 
} 
For $x_f \gg x_f^{\rm AF}$ the coupling or conformal expansion is no longer reliable, 
but one can still analytically explore the phase diagram 
by means of the large $N_f$ expansion \cite{Gracey:1996he,Holdom:2010qs} 
which in turns has proven its worth by discovering the aforementioned UV safe phases 
\cite{Litim:2014uca,Antipin:2017ebo}.

The purpose of this work is to estimate the critical number of flavors $N_f^{\rm cr}$, 
corresponding to the phase boundary between the chirally broken and the IR conformal, 
in $G=SU(N)$, $Sp(2N)$ and $SO(N)$ gauge theories with fermion matter content 
in a single representation $R$. 
In particular, we consider the fundamental (F), adjoint (Adj), two-index symmetric (S2) and antisymmetric (AS), 
and spinorial (S) representations. 
To do this we follow the approach discussed in Ref.~\cite{Kim:2020yvr} 
(see also Ref.~\cite{Appelquist:1998rb} for an earlier work along this direction): 
$N_f^{\rm cr}$ is determined by using the conjectured critical condition 
to the anomalous dimension of a fermion bilinear operator at an IR fixed point, 
$\girpsi=1$ or equivalently $\Girpsi=1$, 
which characterizes the chiral phase transition through the annihilation of infrared and ultraviolet fixed points \cite{Kaplan:2009kr}\footnote{%
Strictly speaking, the critical condition on $\girpsi$ only manifests itself in the large $N$ limit \cite{Gorbenko:2018ncu}. 
We will come back to this issue in \Sec{conformal_phase}. 
} 
and Schwinger-Dyson analysis in the ladder approximation \cite{Appelquist:1988yc,Cohen:1988sq,Appelquist:1998rb}. 
To make our analysis scheme-independent, we employ the conformal expansion for the computation of $\girpsi$ \cite{Ryttov:2016hdp}. 
At finite order in the conformal expansion 
the two critical conditions lead to different results in general, 
and the latter definition is often used because it does not only show better convergence to the known orders \cite{Kim:2020yvr} 
but also reproduces the value of the critical coupling $\alpha^{\rm cr}$ obtained from 
the Scwhinger-Dyson analysis in the ladder approximation \cite{Appelquist:1998rb}. 
As we discuss in \Sec{conformal_phase} in detail, 
the critical condition quadratic in $\girpsi$ is further supported by the fact that 
$|1-\girpsi|$ has a square-root singularity with respect to $N_f-N_f^{\rm cr}$ 
if the IR and UV fixed point merger is concerned. 
For the rest of this paper we reluctantly use the simplified notation $\gir$ for $\girpsi$. 

In this work we also put one step forward by taking account of 
the uncertainties associated with the truncation of the conformal expansions at finite order. 
It is largely unknown whether the conformal expansion is convergent or divergent asymptotic,\footnote{%
The conformal expansion is expected to be free of factorially increasing coefficients 
due to renormalons which 
dominate the large-order behavior of the coupling expansion in QCD-like theories 
\cite{Brodsky:2000cr,Gardi:2001wg}. 
Such a fact results in the better-behaved series expansions 
which have been explicitly shown in the higher-order calculations of $\girpsi$ and $\beta'_{\rm IR}$ 
\cite{Ryttov:2017toz,Ryttov:2017kmx}. 
For instance, the convergence of the coupling expansion (in the $\overline{\rm MS}$-scheme) is challenged from the fact that
the $5$-loop order result for the $SU(3)$ gauge theory coupled to $N_f=12$ fundamental fermions
finds no IR fixed point in contrast to the results at lower-loop orders
\cite{Ryttov:2016ner}, 
while the conformal expansion to the highest order known to us 
shows no explicit evidence of any divergent features \cite{Ryttov:2016asb}. 
Of course, the absence of renormalons is not sufficient to conclude that the conformal series 
is convergent, since other types of factorial growth such as the one related to the multiplicity 
of diagrams could be involved. 
Also, in Ref.~\cite{DiPietro:2020jne} it has been argued that the conformal series would be divergent 
asymptotic if the coupling expansion turns out to be divergent. 
} 
and how far the expansion can reliably be applicable.
Since there has been much evidence that the conformal expansion 
may reach the lower end of the conformal window (e.g., see Refs.~\cite{Ryttov:2016asb,Ryttov:2016hal}), 
we first assume that this is the case. 
We then treat the two possibilities for the asymptotic behavior of conformal expansion, separately. 
In case the expansion is convergent, we employ the Pad\'e approximation 
to approximate the closed forms for the two definitions of the critical condition. 
The difference in the resulting values of $N_f^{\rm cr}$ will be taken 
as the uncertainty of our analysis. 
The best Pad\'e approximants are determined by comparing 
their asymptotic behaviors at sufficiently large values of $N_f$ to the large $N_f$ expansion. 
If the conformal series is assumed to be divergent asymptotic, on the other hand, 
we roughly estimate the uncertainty, 
associated with the truncation at the largest order available up to date, 
by approximating the size of an ambiguity in the perturbative expansion 
which is closely related to the singularity in the Borel plane. 

The paper is organized as follows. 
In \Sec{conformal_phase} 
we discuss the generic infrared properties of non-abelian gauge theory coupled to fermionic matter 
at zero temperature and chemical potential by focusing on the asymptotically free regime. 
In particular, we recall that both the truncated Schwinger-Dyson analysis and the mechanism 
of fixed-point annihilation yield the same critical condition, $\girpsi=1$ or equivalently $\Girpsi=1$, 
characterizing the loss of conformality in the IR. 
In \Sec{conformal_expansion}, we briefly review the conformal expansion of $\girpsi$ defined at an IR fixed point. 
We then describe our strategy to determine the lower edge of the conformal window in a scheme independent way in \Sec{critical_condition}: 
we apply the critical condition, which is responsible for the chiral phase transition, 
to $\girpsi$ computed from the conformal expansion at finite order. 
\Sec{error_estimate} is devoted to estimate the size of systematic effects 
associated with the finite-order perturbative calculations by assuming the different large-order behaviors of the conformal expansion, 
convergent or divergent asymptotic. 
We present our main results on the conformal window of $SO(N),\, SU(N)$ and $Sp(2N)$ gauge theories 
with $N_f$ Dirac (or $N_{Wf}$ Weyl) fermions in various representations, in \Sec{large_n} for the large $N$ limit 
and in Secs.~\ref{sec:cw_su}, \ref{sec:cw_sp} and \ref{sec:cw_so} for finite values of $N$, respectively. 
We critically assess our results by comparing to other analytical methods 
and the most recent nonperturbative lattice results available in the literature. 
Finally, we conclude by summarizing our findings in \Sec{conclusion}. 

\section{Background and methods}
\label{sec:background}

\subsection{Infrared conformal phase in asymptotically free gauge theories}
\label{sec:conformal_phase}

We consider a generic non-abelian gauge theory containing $N_f$ flavors of massless fermionic matter 
in distinct representations $R$ 
of the gauge group $G=SO(N)$, $SU(N)$, and $Sp(2N)$.\footnote{%
Throughout this section $N_f$ denotes the dummy variable for either the Dirac or Weyl flavors 
unless explicitly specified. 
} 
The evolution of the gauge coupling constant $g$ is described by the renormalization group beta function
\beq
\beta(g)=\frac{dg}{dt},
\eeq
where $t={\rm ln}\mu$ with $\mu$ the renormalization scale. 
For a small value of $g$ the RG evolution can be studied by perturbation technique in a reliable way, 
which is equivalent to the Feynman loop expansion. 
After rewriting the coupling constant as $\alpha=g^2/4\pi$ to be positive definite, 
one can write the perturbative beta function as
\beq
\beta(\alpha)=-2\alpha \sum_{\ell=1}^\infty b_\ell \left(\frac{\alpha}{4\pi}\right)^\ell,
\label{eq:pert_beta}
\eeq
where the $\ell$-loop coefficient $b_\ell$ depends on the details of the theory, such as the number of flavors $N_f$, the number of colors $N$, 
the representation $R$, and the gauge group $G$. 

The essential features of the perturbative theory are encoded in the lowest two terms 
which are independent of the renormalization scheme. 
Note that in general the series expansions at finite order in $\alpha$ for $\ell \geq 3$ are scheme-dependent. 
With $N_{f}$ Dirac fermions in the representation $R$ of the gauge group $G$ 
the explicit expressions of $b_1$ \cite{Gross:1973id,Politzer:1973fx} and $b_2$ \cite{Caswell:1974gg} are
\beq
b_1&=&\frac{11}{3}C_2(G)-\frac{4}{3} N_{f} T(R),\label{eq:b1}\\
b_2&=&\frac{34}{3}C_2(G)^2-\frac{4}{3}\left(5 C_2(G)+3 C_2(R)\right) N_f T(R),
\label{eq:b2}
\eeq
where $T(R)$ is the trace normalization factor and $C_2(R)$ is the quadratic Casimir invariant with $C_2(G)=C_2(\rm Adj)$.\footnote{%
The group theoretical invariants are defined as 
${\rm Tr}[T^a_R T^b_R]=T(R)\delta^{ab}$ and $T^a_R T^a_R=C_2(R) I$, 
where the summation runs over $a=1,\cdots,\,d_G$ with $d_G$ the dimension of the gauge group $G$. 
Here, $T^a_R$ are the generators in the representation $R$ of $G$ and 
the group invariants are related by $C_2(R) d_R = T(R) d_G$ with $d_R$ the dimension of the representation $R$. 
Our conventions for the normalization of the generators $T_R$ are 
$T(F)=1/2$ for the fundamental representations of $SU(N)$ and $Sp(2N)$, and $T(F)=1$ for $SO(N)$, 
where the explicit values of other group invariants relevant to this work can be found in the Appendix.~A of Ref.~\cite{Kim:2020yvr}. 
} 
The beta function in \Eq{pert_beta} has a trivial fixed point at $\alpha=0$, a Gaussian fixed point, 
for which the theory is free. 
In the vicinity of this fixed point the coupling constant can be arbitrarily small 
and the behavior of the RG flow is governed by the slope of the beta function, i.e., the sign of $b_1$. 
Consider that we fix the gauge group, the fermion representation, and the number of colors, 
but continuously vary the number of flavors $N_f$ for which only non-negative integer values are physically meaningful. 
For sufficiently small $N_f$ the coefficient $b_1$ has a positive value 
and the coupling constant approaches zero as the momentum scale flows from the IR to the UV, 
indicating that the theory is asymptotically free at high energy. 
If the number of flavors is larger than $N_f^{\rm AF}=11C_2(G)/4T(R)$ or equivalently $b_1<0$, 
on the other hand, the theory loses the asymptotic freedom in the UV, while the IR theory is trivial. 
The focus of our interest is in the asymptotically free theory and thus we restrict our attention to $N_f < N_f^{\rm AF}$. 

The $2$-loop results further divide the asymptotically free region into two nontrivial phases 
whose IR behaviors are distinct from each other. 
If the number of flavors is sufficiently small such that $b_2>0$, including the extreme case of the pure Yang-Mills ($N_f=0$), 
from the UV to the IR the coupling runs to infinity and the theory is expected to confine by developing a dynamical scale. 
In the presence of fermionic matter the global (flavor) symmetry is also expected to be broken due to the nonzero fermion condensate. 
From the fact that a negative value of $b_2$ and an arbitrarily small positive value of $b_1$ are realized if $N_f$ is just below $N_f^{\rm AF}$, 
on the other hand, one finds a coupling constant satisfying $\beta(\alpha_{\rm BZ})=0$ at $\alpha_{\rm BZ}=-4\pi b_1/b_2 \ll 1$ in a reliable manner 
within the perturbation theory \cite{Caswell:1974gg,Banks:1981nn}. 
The corresponding BZ fixed point, named after Banks-Zaks, 
suggests the existence of interacting IR conformal theories (even beyond the weak coupling regime) 
with certain numbers of flavors ranged over $N_f^{\rm cr}< N_f < N_f^{\rm AF}$. 
Such an interval in $N_f$ is commonly called the {\it conformal window} (CW). 
The conformal phase near the upper bound can systematically be studied by the perturbative analysis as discussed above. 
However, it is difficult to investigate the phase near the lower bound, 
because in general $\alpha_{\rm BZ}$ grows to a large value as $N_f$ decreases 
and thus the coupling expansion would be no longer reliable. 
In this region, nonperturbative effects are also expected to be sizable in the IR. 
It is even a nontrivial task to determine the value of $N_f^{\rm cr}$. 

In Ref.~\cite{Kaplan:2009kr}, it has been argued that the underlying mechanisms responsible for the loss of conformality could generally be classified 
by the following three criteria from the RG point of view: 
(a) the coupling at an IR fixed point, $\alpha_{\rm IR}$, goes to zero, (b) $\alpha_{\rm IR}$ runs off to infinity, 
or (c) an IR fixed point merges with a counterpart UV fixed point. (See also Ref.~\cite{Gorbenko:2018ncu}.) 
The transition between asymptotically free and unfree phases at $N_f^{\rm AF}$, the upper bound of CW, 
belongs to scenario (a), where the BZ fixed point annihilates with the Gaussian fixed point at zero coupling. 
Similarly, the lower end of the conformal window might be determined from scenario (b) using the $2$-loop results, 
i.e. $b_2=0$ such that $\alpha_{\rm BZ}\rightarrow \infty$. 
However, such a naive estimation is limited by the breakdown of the perturbative expansion in the first place. 
Nevertheless, 
we note that mechanism (b) successfully describes the conformal transition at the lower end of the conformal window 
in $\mathcal{N}=1$ supersymmetric QCD (SQCD) 
through the electro-magnetic (Seiberg) duality \cite{Seiberg:1994pq,Intriligator:1995au}, 
where the loss of conformality in the dual magnetic theory is described by scenario (a) in a weak coupling regime. 
In other words, both the mechanisms of (a) and (b) are equivalent to each other up to the reparametrization of the coupling constant, 
and could be categorized by a more generic scenario of that the fixed point merges with that of a free field theory \cite{Gorbenko:2018ncu}. 
Since no chiral symmetry breaking happens even outside the conformal window in this scenario, 
it cannot describe the chiral phase transition. 

The last scenario was realized in the exemplified cases of certain nonrelativistic and relativistic quantum theories, 
and conjectured to explain the loss of conformality at the lower end of the conformal window in four-dimensional (4d) 
nonsupersymmetric gauge theories in the large $N$ limit \cite{Kaplan:2009kr}.\footnote{
See also Ref.~\cite{Gies:2005as}, where the merger of two fixed points, as the resultant of the functional analysis, 
was suggested to explain the chiral phase transition in QCD with many fundamental flavors. 
The authors also estimated the critical number of flavors $N_f^{\rm cr}$ including theoretical errors. 
}~\footnote{
Although mechanism (c) still remains a conjecture for the chiral phase transition because no rigorous proof exist, 
it is encouraging that UV complete 4d gauge models explicitly realizing the scenario of merging two fixed points 
were found in the weakly coupled regime \cite{Benini:2019dfy}.
} 
The UV ($+$) and IR ($-$) fixed points correspond to the solutions
of the suggested RG equation, 
$\beta(g;\alpha)\equiv -(\alpha-\alpha^{\rm cr})-(g-g^{\rm cr})^2=0$ with $(\alpha^{\rm cr}-\alpha)>0$, 
i.e. $g_\pm=g^{\rm cr}\pm\sqrt{\alpha^{\rm cr}-\alpha}$.\footnote{
For $\alpha > \alpha^{\rm cr}$ the solutions become complex and the conformal symmetry is expected to be broken. 
If $|\alpha-\alpha^{\rm cr}| \ll 1$, the RG beta function yields that the coupling $g$ is {\it walking}, instead of running, near $g=g_{\rm IR}$, 
and the IR scale would be largely separated from the UV scale, i.e. $\Lambda_{\rm IR}/\Lambda_{\rm UV} = e^{-\pi/\sqrt{\alpha-\alpha^{\rm cr}}}$ \cite{Kaplan:2009kr}. 
It has been suggested that such theories, exhibiting walking behavior, could systematically be studied 
by complex conformal field theories (CFTs), where the real RG flow goes through 
the pair of complex fixed points having small imaginary parts \cite{Gorbenko:2018ncu}.
} 
If we consider non-abelian gauge theories coupled to fermionic matter in $d=4$ space-time dimension, 
the interpretation of the couplings is as follows: 
$g$ is the dimensionless running coupling associated with a certain scalar operator $\mathcal{O}_g$ 
which becomes marginal at the chiral phase transition, 
$\alpha$ is the gauge coupling at an approximate fixed point 
which is (almost) constant and thus treated as an external parameter, 
and $g^{\rm cr}$ and $\alpha^{\rm cr}$ are the critical couplings in the onset of the loss of conformality. 
The coupling $\alpha$ can be tuned by varying the number of flavors $N_f$ (or $x_f=N_f/N_c$ in the Veneziano limit). 
We note that the beta function $\beta(g;\alpha)$ only provides a good approximation to the RG evolution near the chiral phase transition, 
i.e., $|\alpha-\alpha^{\rm cr}| \ll 1$ and $|g-g^{\rm cr}| \ll 1$. 
At the UV and IR fixed points the RG equation yields that 
$\Delta_{\mathcal{O}_g,\pm}\equiv 4+\beta'(g_\pm)=4\mp 2\sqrt{\alpha^{\rm cr}-\alpha}$ with $\Delta_{\mathcal{O}_g}$ the scaling dimension of $\mathcal{O}_g$. 
Approaching the lower end of conformal window the dimension $\Delta_{\mathcal{O}_g}$, as well as the coupling $g$, 
has a square-root singularity in the vicinity of $\alpha^{\rm cr}$, i.e., $|g-g^{\rm cr}| \sim |\Delta_{\mathcal{O}_g}-4| \sim \sqrt{\alpha^{\rm cr}-\alpha}$. 
In terms of the number of flavors $N_f$ (or $x_f$), which basically controls how $\alpha$ approaches $\alpha^{\rm cr}$, 
one can find the analogous square-root singularity of $\Delta_{\mathcal{O}_g}$, 
provided that the $\alpha$ is an analytic function of $N_f$ at $N_f^{\rm cr}$, as 
\beq
|\Delta_{\mathcal{O}_g}-4| \sim \sqrt{N_f-N_f^{\rm cr}}.
\label{eq:gamma_O}
\eeq
Such a square-root behavior has been advocated and emphasized in Ref.~\cite{Gukov:2016tnp}. 

The most natural candidate for $\mathcal{O}_g$ in the large $N$ limit 
would be the chirally symmetric four-fermion operators of double-trace form 
whose mass dimension is $6$, e.g. $(\bar{\psi}\gamma^\mu\psi)^2$ \cite{Gubser:2002vv}. 
The large $N$ factorization yields that the dimension of the fermion bilinear is 
exactly half of that of the double-trace four-fermion operator at infinite $N$, 
and the above discussion would lead to the critical condition to the anomalous dimension, $\gpsi=1$ \cite{Kaplan:2009kr}. 
At finite but large $N$, of course,  it is believed to receive finite $N$ corrections as $\gpsi
=1+\mathcal{O}(1/N)$ (for instance, see Sec.~D of Ref.~\cite{Gorbenko:2018ncu}).\footnote{
At finite $N$ the marginal operator at $N_f^{\rm cr}$ could also take the form of other than double-trace or a mixture. 
For instance, see Ref.~\cite{Benini:2019dfy} for the discussion about the mixture of single- and double-trace operators 
at finite $N$ in the context of weakly coupled and UV complete gauge models realizing the scenario of fixed point merger. 
} 
Again, in the vicinity of the chiral phase transition it is expected that $\gpsi$ has a square-root singularity 
by approaching the lower end of the conformal window from above (up to $1/N$ corrections),
\beq
|1-\gpsi| \sim \sqrt{\alpha^{\rm cr}-\alpha} \sim \sqrt{N_f-N_f^{\rm cr}}.
\label{eq:g_square_root}
\eeq
As we discuss in \Sec{conformal_expansion}, we compute $\gpsi$ at the IR fixed point using the Banks-Zaks conformal expansion 
and thus may not capture such a nonanalytic behavior properly. 
To take the full advantage of the conformal expansion, we therefore define an alternative form of the critical condition, 
\beq
\girpsi (2-\girpsi) = 1,
\label{eq:critical_condition_2}
\eeq
which is equivalent to $\girpsi=1$, but the left-hand side is a linear function of $N_f-N_f^{\rm cr}$ away from the lower end of the conformal window. 

The critical condition on $\girpsi$ has also been discussed when one studies the chiral phase transition 
by solving the Schwinger-Dyson (SD) gap equation in the ladder approximation. 
In particular, the suggested form of the critical condition was found in Ref.~\cite{Appelquist:1988yc}, 
$\Girpsi=\alpha/\alpha^{\rm cr}=1$. 
Surprisingly, this critical condition is exactly same with the one discussed above up to $\mathcal{O}(1/N)$ corrections, e.g., \Eq{critical_condition_2}. 
As argued in Ref.~\cite{Cohen:1988sq}, furthermore, this critical condition is believed to persist beyond the ladder approximation. 
In terms of the gauge coupling the truncated SD equation yields the critical condition, $\alpha_{\rm IR}=\alpha^{\rm cr}$ with $\alpha^{\rm cr}=\pi/3C_2(R)$. 
The naive and traditional way to determine $N_f^{\rm cr}$ using the SD analysis is 
to calculate $\alpha_{\rm IR}$ from the $2$-loop beta function and match it to $\alpha^{\rm cr}$. 
However, if we want to proceed the analysis beyond the $2$-loop, we cannot avoid renormalization-scheme dependence and the critical condition becomes ambiguous. 
We therefore use the anomalous dimension $\girpsi$ instead, which is physical and thus scheme-independent, to estimate the critical number of flavors $N_f^{\rm cr}$ 
by employing the aforementioned critical condition. 
This condition satisfies the unitarity condition by construction, 
$\girpsi \leq 2$ \cite{Mack:1975je}. 
We note that in the case of SQCD the unitarity condition is the same with the onset of the loss of conformality 
and has often been used to determine the conformal window even for nonsupersymmetric theories. 
However, these two conditions could largely be different in general, 
because the underlying mechanism for the loss of conformality is expected to depend on the details of the theory as discussed above. 

\subsection{Conformal expansion for the anomalous dimension $\girpsi$}
\label{sec:conformal_expansion}

One of the consequences of the perturbative BZ fixed point 
is that the IR coupling can be expanded in terms of 
the distance from $N_f^{\rm AF}$, $\Delta_{N_f}\equiv N_f^{\rm AF}-N_f^{\rm IR}$, 
i.e., the Banks-Zaks conformal expansion \cite{Banks:1981nn}, 
\begin{equation}
\frac{\alpha_{\rm IR}}{4\pi}=\sum_{j=1}^{\infty}a_{j} (\Delta_{N_f})^{j}\,,
\end{equation}
where the coefficients $a_j$ are independent of $N_f$. 
The leading order term is solely determined from the two-loop results as 
\beq
\alpha_{\rm IR}
=4\pi a_1 \Delta_{N_f} +\mathcal{O}(\Delta_{N_f}^2),
\label{eq:cx_alpha}
\eeq
with
\beq
a_1=\left.\frac{1}{b_2}\frac{\partial b_1}{\partial N_f}\right|_{N_f=N_f^{\rm AF}}=\frac{4 T(R)}{3 C_2(G) (7C_2(G)+11 C_2(R))}.
\eeq 
Similarly, the $j$th order coefficient $a_j$ can be determined from 
a power series solution to $\beta(\alpha)=0$ with $\beta(\alpha)$ in \Eq{pert_beta} truncated at the $(j+1)$th order. 

The most notable feature of the conformal expansion is that the series coefficients of the expansion for a physical observable 
are {\it universal}, in the sense that they are independent on the renormalization scheme order by order~\cite{Ryttov:2016hdp}. 
Such a fact can be understood on general grounds, because the expansion parameter $\Delta_{N_f}$, 
defined through the scheme-independent $2$-loop beta function, is a well-defined physical quantity. 
The conformal expansion relevant to us is the one for the anomalous dimension of a fermion bilinear operator
\beq
\label{eqn:cx_gamma}
\girpsi(\Delta_{ N_{f}})=\sum_{j=1}^{\infty} c_{j}(\Delta_{N_f})^{j}.
\eeq
The coefficients $c_j$ are determined by combining the results of the coupling expansion of $\girpsi(\alpha)$ 
at the $j$th order and $\beta(\alpha)$ at the $(j+1)$th order, respectively. 
As mentioned above, the conformal expansion is scheme-independent at a given order and 
does not require any information from higher-order terms. 
This is a somewhat distinctive feature compared to other expansions, alternative to the coupling expansion, such as the large-$N_f$ expansion 
for which all orders in $\alpha$ are necessary to compute the coefficient at each order in $1/N_f$. 
The recent computations of the perturbative beta function at the $5$th order in the gauge coupling 
\cite{Baikov:2016tgj,Herzog:2017ohr,Luthe:2017ttg,Chetyrkin:2017bjc}, 
along with the results for the anomalous dimension at the $4$th order 
\cite{Chetyrkin:1997dh,Vermaseren:1997fq}, 
within the modified minimal subtraction ($\overline{\rm MS}$) scheme 
enable to determine the coefficients $c_\ell$ to the $4$th order in $\Delta_{N_f}$, 
where the explicit results in terms of group invariants for fermions transforming according to the representation $R$ 
of a generic gauge group $G$ are presented in Ref.~\cite{Ryttov:2017kmx}. 

In general, besides the scheme-independence, the conformal expansion of $\girpsi$ better behaves 
compared to the coupling expansion. 
For instance, we refer the reader to the results in Tables.~1-5 of Ref.~\cite{Ryttov:2017kmx}, 
where the resulting values of $\girpsi$, computed using both the conformal and coupling expansion 
up to $4$th order for $SU(N)$ gauge theories coupled to $N_f$ Dirac fermions 
in the fundamental, adjoint, two-index symmetric and antisymmetric representations, are present. 
First of all, at fixed $N_f$, 
$\girpsi$ monotonically increases with the order of $\Delta_{N_f}$ for all the theories considered, 
indicating that the coefficients are all positive, 
while those in the coupling expansion are not. 
Furthermore, the difference of $\girpsi$ between the adjacent orders of $\Delta_{N_f}^j$ and $\Delta_{N_f}^{(j+1)}$ 
typically decreases if $j$ increases, 
except for $j=2$ and $3$ in a few theories with small values of $N_f$ ($x_f$). 
Interestingly, these exceptional cases consistently reside in the broken phase 
just outside the conformal window estimated in this work. 
If the conformal expansion turns out to be a convergent series, such an agreement would indicate that the radius of convergence 
is closely related with the phase boundary at which the loss of conformality occurs. 
As we have no good understanding of this interesting observation at the moment, however, 
it should not be generalized to a generic feature of the chiral phase transition in 
nonsupersymmetric theories without further investigation. 
Last but not the least, $\girpsi$ at each order in the conformal expansion also monotonically increases 
with $\Delta_{N_f}$ due to the positive coefficients. 
Accordingly, with a few lowest coefficients the perturbative calculations of $\girpsi$ 
are well stretched to the very small $N_f$. 
However, we note that the results at small $N_f$ should be taken with some care, 
because the conformal expansion only makes sense when an IR fixed point exists, 
where its existence is not known {\it a priori}.

\subsection{Determination of the lower edge of the conformal window}
\label{sec:critical_condition}

Following the discussions we had in \Sec{conformal_phase},  
we assume that the chiral phase transition in nonsupersymmetric gauge theories occurs through  mechanism (c) rather than (b), 
i.e. an IR fixed point disappears by annihilating a UV fixed point. 
Furthermore, we borrow the large $N$ argument 
and assume that the chiral phase transition is triggered by the four-fermion operators of double-trace form. 
To determine the critical number of flavors $N_f^{\rm cr}$, 
corresponding to the lower edge of conformal window, we therefore 
adopt the following critical condition to the anomalous dimension of a fermion bilinear,
\beq
\girpsi=1~~{\rm or~equivalently}~~\Girpsi=1.
\label{eq:critical_condition}
\eeq
As discussed in \Sec{conformal_phase}, we expect to have finite $N$ corrections for the gauge theories apart from the infinite $N$ limit. 
We will discuss the implications of such finite $N$ effects on the resulting values of $N_f^{\rm cr}$ in \Sec{cw}. 

In an earlier work along this direction \cite{Appelquist:1999hr}, the coupling expansion including higher order terms 
in the $\overline{\rm MS}$ scheme was used to compute $\girpsi(\alpha_{\rm IR})$ and $\alpha_{\rm IR}$. 
Furthermore, the authors employed the critical condition $\Girpsi=1$, not $\girpsi=1$, 
because the $1$-loop result turned out to be identical to the critical condition on $\alpha$ in the traditional Schwinger-Dyson analysis. 
In this work we instead use the Banks-Zaks conformal expansion for the computation of $\girpsi$, 
because it shows better behavior as discussed in the previous section. 
More importantly both forms of the critical condition can be expanded order by order in a scheme independent way, 
so be the conformal window. 
Comparisons between the conformal and coupling expansions for the determination of $N_f^{\rm cr}$ 
in $SU(3)$ gauge theories coupled to $N_f$ fundamental Dirac fermions 
are found in Ref.~\cite{Kim:2020yvr}. 

Two equivalent critical conditions in \Eq{critical_condition} should be identical to each other 
if all orders of the conformal expansion are considered. 
If the left-hand sides of those equations are truncated at the finite order $n$, 
however, it leads to two different critical conditions. 
Accordingly, the resulting values of $N_f^{\rm cr}$ are different in general. 
To be explicit, we first define the finite-order critical condition of the former
\beq
\gamma_{\rm IR}^{(n)}(\Delta_{N_f})=\sum_{j=1}^{n}{k_{j}\left(\Delta_{N_f}\right)^{j}} \equiv 1, 
\label{eq:cc1_finite}
\eeq
where the coefficients $k_j$ are known to the $4$th order \cite{Ryttov:2017kmx}. 
Similarly, the latter critical condition at each order $n$ can be written as
\beq
\left[\gamma_{\rm IR}(2-\gamma_{\rm IR})\right]^{(n)}(\Delta_{N_f})=\sum_{j=1}^{n}{\kappa_{j}\left(\Delta_{N_f}\right)^{j}}\equiv 1. 
\label{eq:cc2_finite}
\eeq
The coefficients $\kappa_j$ are related to $k_j$ as
\beq
\kappa_1=2k_1,~\kappa_2=2k_2-k_1^2,~
~\kappa_3=2k_3-k_1 k_2,~
\kappa_4=2k_4-2k_1 k_3-k_2^2,~\cdots.
\label{eq:coeff_cc2}
\eeq

\begin{figure}
\begin{center}
\includegraphics[width=0.59\textwidth]{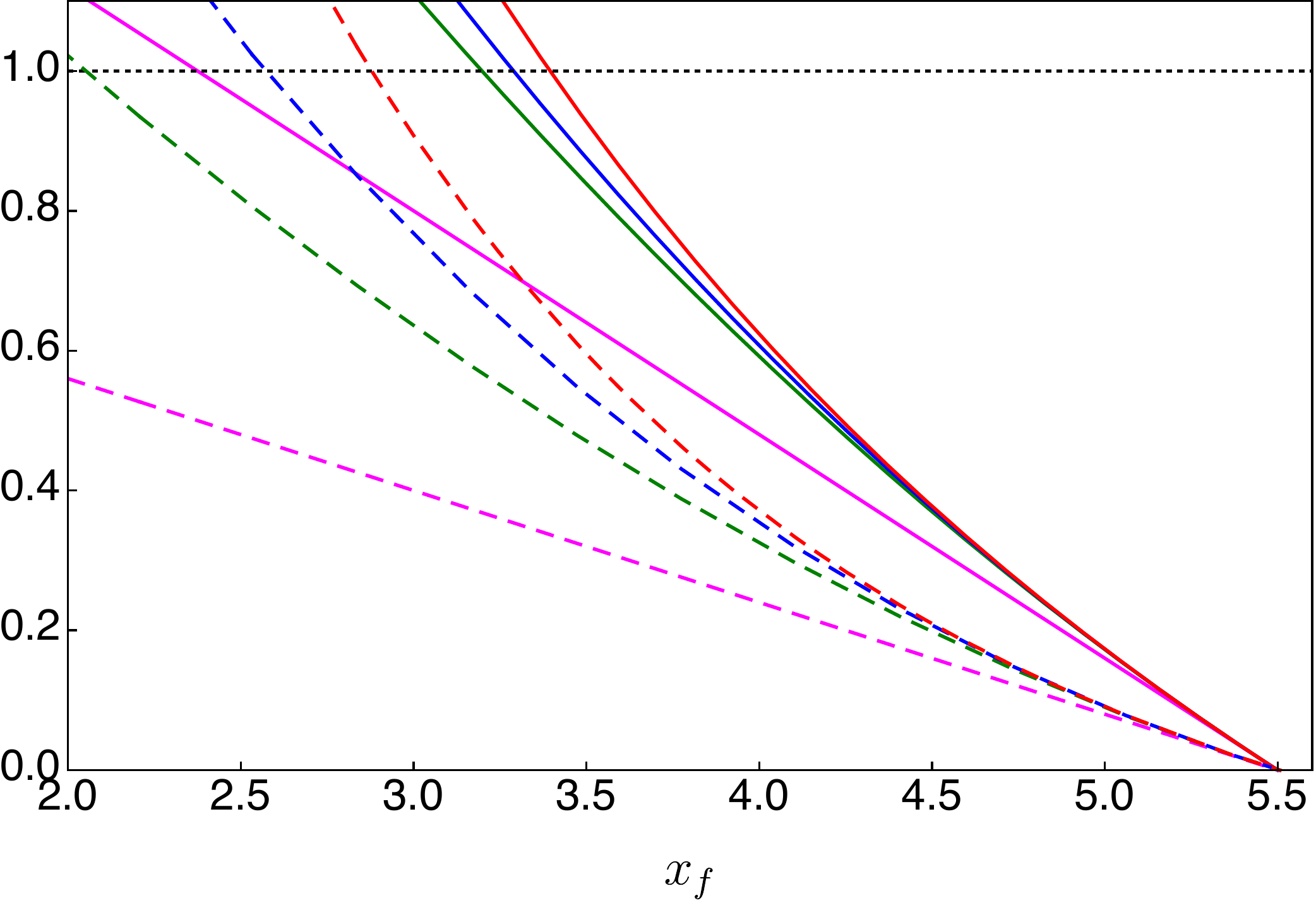}
\caption{%
Conformal expansions of $\girpsi$ (dashed lines) and $\Girpsi$ (solid lines) at finite order in $\Delta_{x_f}=11/2-x_f$
for $SU(N)$ gauge theories coupled to the $N_f$ Dirac flavors of fundamental fermions in the Veneziano large $N$ limit 
with fixed $x_f=N_f/N$. 
Purple, green, blue, and red colors denote the results obtained by truncating the series expansions at $n=1,\,2,\,3$ and $4$, respectively. 
The black dotted line corresponds to the critical condition for the loss of conformality. 
}
\label{fig:pert_gamma}
\end{center}
\end{figure}

To illustrate the typical behavior of the left-hand sides of \Eq{cc1_finite} and \Eq{cc2_finite}, 
we consider $SU(N)$ gauge theories coupled to $N_f$ fundamental Dirac fermions in the Veneziano limit, 
i.e. $N\rightarrow \infty$ and $N_f\rightarrow \infty$ with the ratio $x_f=N_f/N$ fixed. 
We define $\Delta_{x_f}=x_f^{\rm AF}-x_f$ with $x_f^{\rm AF}=11/2$. 
In \Fig{pert_gamma}, we show the results for $x_f\leq x_f^{\rm AF}$. 
As discussed in the previous section, $\gir^{(n)}(\Delta_{x_f})$ monotonically increases as we go to the higher order in $\Delta_{x_f}$ 
over the whole range of $x_f$ considered, so does $\left[\Gir\right]^{(n)}(\Delta_{x_f})$. 
We also observe that the latter receives smaller corrections from higher order terms along the black dotted line, 
corresponding to the critical condition, 
which can be understood as follows. 
First of all, the monotonic increment of $\gir$ with $n$ implies the positiveness of the coefficients $c_i$, 
which in turn results in $\kappa_{2}/\kappa_1 < k_{2}/k_1$ as seen in \Eq{coeff_cc2}, 
i.e., for a given value of $\Delta_{x_f}$ the ratio between the second and first terms of $\Gir$ 
is smaller than that of $\gir$. 
We note that such an inequality cannot always be true for higher order coefficients. 
Secondly, the leading-order result of $\Gir$ starts by twice larger than that of $\gir$. 
Combined with the positive coefficients, it leads us to find higher-order results at smaller $\Delta_{x_f}$, 
which allows us to be in the better controlled regime of the perturbative series expansion. 

As we approach the unity, therefore, we find that the conformal expansion of $\Gir$ 
provides better performance to estimate $x_f^{\rm cr}$. 
As shown in the left panel of \Fig{dxf}, in particular, 
the differences in the resulting values of $\Delta_{x_f}$ 
(or equivalently $x_f^{\rm cr}$) among $n=2,\,3,$ and $4$th orders (red circle)
are within  $10\%$ level, while those obtained from the conformal expansion of $\gir=1$ (blue circle) are about $20\sim 40 \%$. 
Nevertheless, we find that red and blue circles approach to each other as we go to the higher order in $\Delta_{x_f}$, 
which is consistent with our expectation. 
Similar conclusions are drawn for the other theories considered in this work. 
Therefore, we use the critical number of flavors $N_f^{\rm cr}$ determined from the finite-order critical condition 
defined in \Eq{cc2_finite} with $n=4$ 
as our best estimate for the lower end of the conformal window. 

We want to close this section by commenting on the different behaviors of $\girpsi$ and $\Girpsi$ in the vicinity of the chiral phase transition 
envisioned by the mechanism of fixed point annihilation discussed in \Sec{conformal_expansion}. 
As indicated by Eqs.~\ref{eq:g_square_root} and \ref{eq:critical_condition_2}, 
$\Girpsi$ would linearly approach the unity as we decreases $x_f$ within the conformal window, 
while $\girpsi$ would approach it with a square-root singularity.  
Although it is not possible to correctly describe the expected singular behavior of $\girpsi$ with the series expansion, 
the results obtained by the first few low-lying terms are indeed 
consistent with the qualitative picture as shown in \Fig{pert_gamma}. 
Such a fact further supports our conclusion that the finite-order critical condition in \Eq{cc2_finite} is more appropriate to determine the lower end of the conformal window 
with a better accuracy.  

\subsection{Error estimates}
\label{sec:error_estimate}

Aside from higher order corrections to the critical condition, 
our approach discussed in the previous section 
could suffer from various unknown systematic effects like as in other analytical methods.
Let us first discuss the nonperturbative effects implied by a certain type of divergence of perturbative expansions. 
The best known example might be the IR renormalon in QCD-like theories which are asymptotically free in the UV 
but confining in the IR. (See Ref.~\cite{Beneke:1998ui} for a classical review.)
The conformal expansion of the mass anomalous dimension, defined at an IR fixed point, is expected to have no IR renormalons by construction \cite{Gardi:2001wg}. 
Other types of divergence, for instance, sourced by instanton anti-instanton configurations, 
may appear in the expansion, but we expect that these are not as severe as that of renormalons. 
Below, we account for such nonperturbative effects, if exist, by investigating the coefficients of the conformal expansion 
and estimate the errors from there. 

We start by assuming that the conformal expansion is a divergent asymptotic series. 
If this is the case, the perturbative series expansion only provides an approximate description at best, 
and we expect to have the best accuracy to the approximant when it is truncated at a certain optimal order $k_{\rm opt}$. 
To see this, we consider a divergent series which is asymptotic to $f(z)$ in a domain $\mathcal{C}$ in a complex $z$ plane
\beq
\sum_{j=0}^{J} c_j z^j. 
\label{eq:fz}
\eeq
Then, for all $z$ in $\mathcal{C}$ there exist numbers $A_J$ such that the error is bounded as
\beq
\left| f(z)-\sum_{j=0}^{J-1} c_j z^j \right| < A_J |z|^J.
\label{eq:fz_bound}
\eeq
If it is assumed that the coefficients factorially increase for $j \gg 1$, $c_j\sim j! \,t_0^{-j}$, 
one might find that $A_J \sim A_0 J! \, t_0^{-J}$ with a constant $A_0$.
The bound will decrease in the beginning, but will eventually diverge as $J\rightarrow \infty$. 
The minimum value of this bound could be identified by the optimal order of $j_{\rm opt}(z)\simeq |t_0|/z$ for which the best accuracy is achieved. 
Roughly speaking, this is the same order at which the subsequent higher-order term becomes larger. 
For this optimal truncation one can find the truncation error of order
\beq
\delta(z) \sim e^{-\frac{|t_0|}{z}}\sim e^{-j_{\rm opt}}.
\label{eq:delta}
\eeq

It might also be instructive to understand the ambiguity of the asymptotically divergent series 
using the standard Borel summation technique. 
The Borel transformation of $f(z)$ is defined as
\beq
\mathcal{B}f(t)=\sum_{j=0}^{\infty} \frac{c_j}{j!} t^j, 
\label{eq:bfz}
\eeq
and its inverse is
\beq
f_B(z)=\int_0^\infty dt e^{-t} \mathcal{B}f(t\,z).
\eeq
With the factorally growing coefficients $c_j$, as before, 
one finds that $\mathcal{B}f(t)$ has a pole at $t=|t_0|/z$ on the positive real axis. 
According to the inverse Borel transformation, 
this singularity in the Borel plane yields the same result in \Eq{delta}. 

Currently, the coefficients of the conformal expansion for $\gir$ are known to the $4$th order in $\Delta_{N_f}$. 
With this limited number of coefficients the search for the optimal truncation discussed above 
should only be understood in a practical sense. 
To demonstrate this, again, consider the Veneziano limit of $SU(N)$ gauge theories with $N_f$ fundamental fermions. 
Since we determine the best estimate of $x_f^{\rm cr}$ using the critical condition as in \Eq{cc2_finite}, 
we focus on the conformal series of $\Gir(\Delta_{x_f})$. 
We first assume that the coefficients $\kappa_j$ exponentially grows in $j$ as for $c_j$ in \Eq{fz_bound} 
and determine $t_0$ at a given order $j$ by $t_0 = j\,\kappa_{(j-1)}/ \kappa_{j}$. 
We obtain $t_0 \simeq 13,\, 32$ and $6$ for $j=2,\,3$ and $4$, respectively. 
As expected, $t_0$ largely varies for these small integer values of $j$. 
Assuming that the highest order result well approximates the large order behavior, 
we read off that  $j_{\rm opt} \simeq t_0/\Delta_{x_f}^{\rm cr} \sim 3$ 
near the lower end of the conformal window with $\Delta_{x_f}^{\rm cr}\sim 2$ as shown in \Fig{pert_gamma}. 
This result indicates that the $4$th order truncation used for the determination of $x_f^{\rm cr}$ 
is roughly the optimal one. 
And we find that the truncation error to $\Gir(\Delta_{x_f^{\rm cr}})$ is simply given by $e^{-j_{\rm opt}}$, 
which might be translated into the critical condition in \Eq{cc2_finite} as
\beq
\left[\Gir\right]^{(n=4)}(\Delta_{N_f}^{\rm cr}) = 1 \mp {\rm exp}\left[{-j_{\rm opt}}\right]. 
\label{eq:te}
\eeq
By solving this equation, one could estimate the uncertainty in $N_f^{\rm cr}$. 

In general, however, we find that $j_{\rm opt}$ is substantially dependent on $G$, $N$ and $R$. 
In particular, in the $SU(N)$ and $Sp(2N)$ theories coupled to fundamental fermions with any values of $N$, as well as some other theories at small $N$, 
the typical values of $j_{\rm opt}$ 
are smaller than the truncation order of $n=3$ at the lower end of the conformal window 
with $\Delta_{N_f}^{\rm cr}=N_f^{\rm AF}-N_f^{\rm cr}$\footnote{
Note that for the comparison to $j_{\rm opt}$ either $n=3$ or $4$ could be used since we determined $j_{\rm opt}$ from the ratio of $\kappa_3$ and $\kappa_4$. 
To make our estimation of the truncation error to be conservative, we take $n=3$.}: 
we can still use \Eq{te} to estimate the uncertainty to $N_f^{\rm cr}$. 
In the other theories considered in this work, on the other hand, we find that $j_{\rm opt}$ is much larger than $3$,
indicating that the optimal truncation is expected to be at far beyond 
the currently available highest order of the conformal expansion, 
and are not able to use the $j_{\rm opt}$ to estimate the truncation error. 
In these cases, practically, we might instead consider the actual truncation order to estimate the truncation error $\sim e^{-3}$. 
To take account of such a large variance in $j_{\rm opt}$ 
we therefore estimate the uncertainty to our determination of $N_f^{\rm cr}$ 
in a conservative and practical way, by generalizing the discussion near \Eq{te}, as
\beq
\delta_1^{\pm} = N_f^{{\rm cr},\,\pm} - N_f^{\rm cr},
\label{eq:delta_1}
\eeq
where $N_f^{{\rm cr},\,\pm}$ are solutions to
\beq
\left[\Gir\right]^{(n=4)}(\Delta_{N_f}^{\rm cr}) = 1 \mp {\rm exp}\left[{-j_{\rm min}}\right], 
\label{eq:truncation_error}
\eeq
with $j_{\rm min}={\rm Min}\left[3,\,t_0/\Delta_{N_f}^{\rm cr}\right]$ and $t_0=4 \kappa_3/\kappa_4$.

\begin{figure}
\begin{center}
\includegraphics[width=0.45\textwidth]{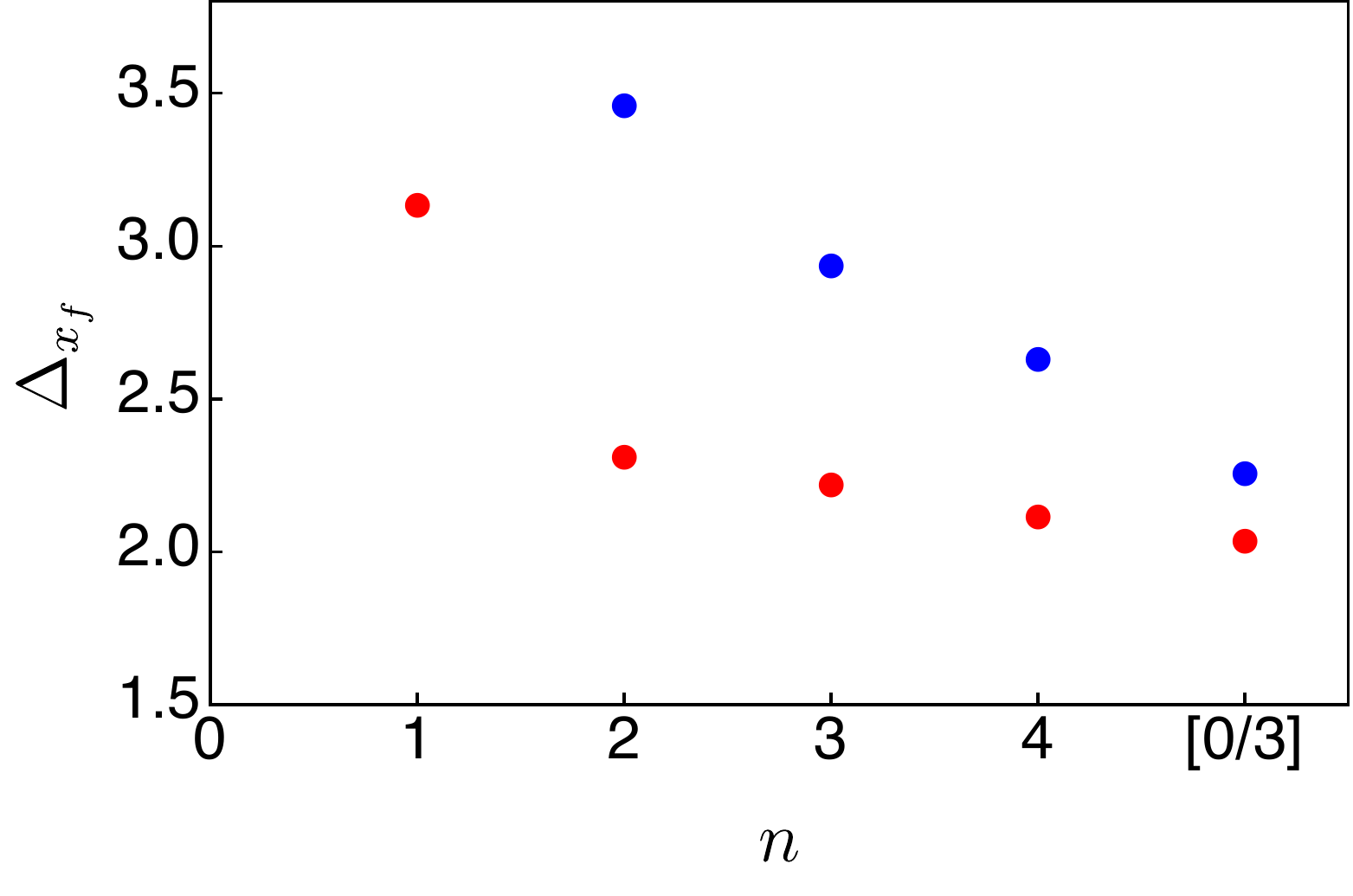}
\includegraphics[width=0.45\textwidth]{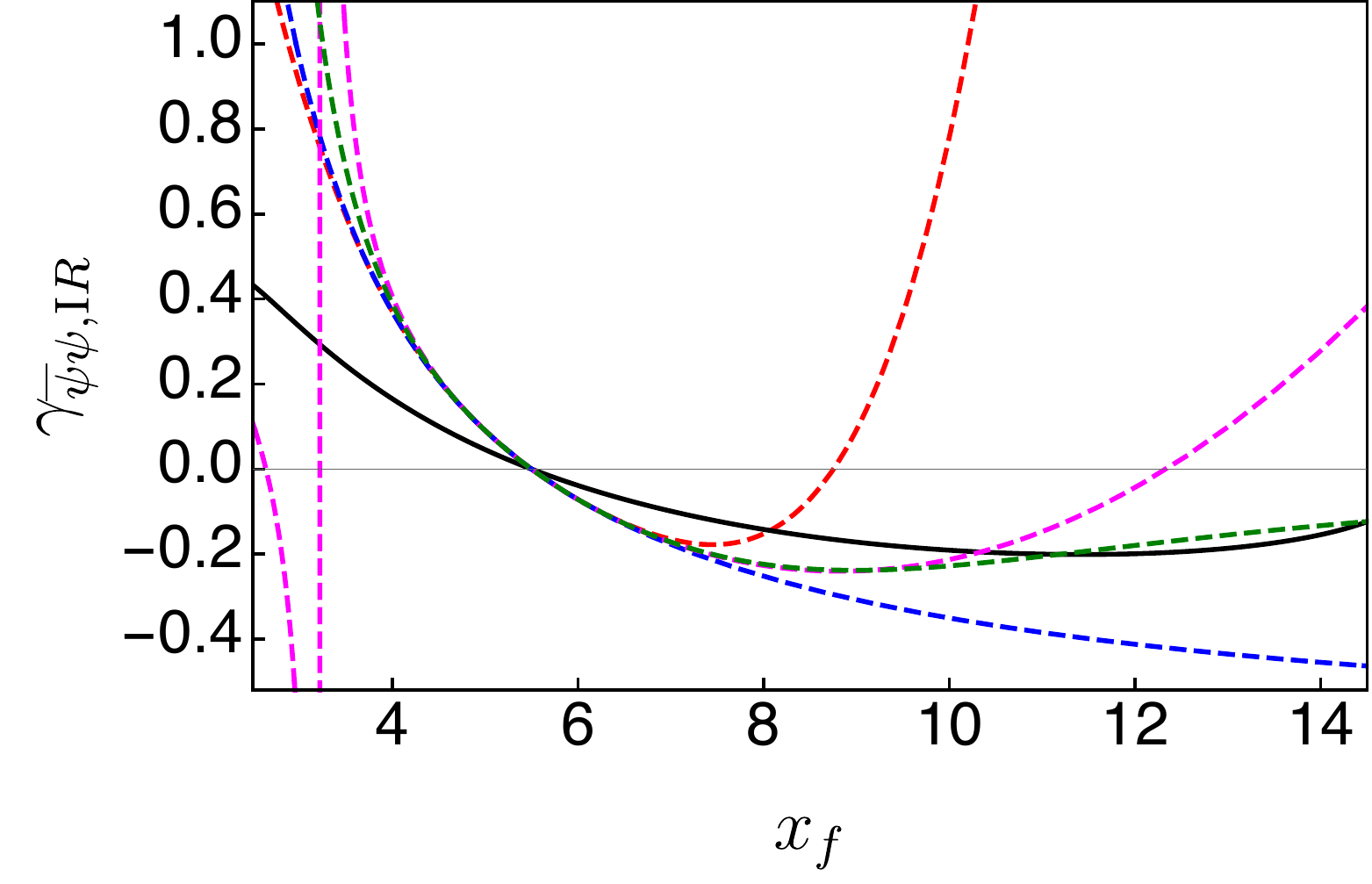}
\caption{%
In the left panel, we present the resulting values of $\Delta_{x_f}=x_f^{\rm AF}-x_f^{\rm cr}$ 
obtained from the finite-order critical conditions in Eqs.~\ref{eq:cc1_finite} and \ref{eq:cc2_finite} truncated at $n$th order, 
denoted by blue and red dots, respectively. 
In the same figure, we also present the results obtained by using the $[0/3]$ Pad\'e approximants 
for the critical conditions in \Eq{critical_condition}. 
In the right panel, we show the results of the anomalous dimension of a fermion bilinear 
at the BZ fixed point in the vicinity of which asymptotic freedom is lost, $x_f^{\rm AF}=5.5$. 
Red dashed line represents for the conformal expansion truncated at the $4$th order, 
black solid line for the leading-order large $N_f$ expansion, 
and purple, blue and green dashed lines for $[2/1]$, $[1/2]$ and $[0/3]$ Pad\'e approximants. 
}
\label{fig:dxf}
\end{center}
\end{figure}

As discussed in \Sec{critical_condition}, the finite-order critical conditions in Eqs.~\ref{eq:cc1_finite} and \ref{eq:cc2_finite} 
generally result in two different values of $N_f^{\rm cr}$, 
and the difference could be taken for the systematic error of our analysis. 
If it is assumed that the corresponding conformal series expansions are convergent, 
one might do a better job by first approximating the closed forms of $\gir$ and $\Gir$. 
To do this, we employ the Pad\'e approximation method (for a review see Ref.~\cite{Baker}). 
The underlying assumption is that $\gir$ is an analytic function of $\Delta_{N_f}$ 
and its Taylor series expansion at $\Delta_{N_f}=0$ is convergent with a finite radius in the complex plane. 
Provided that the series coefficients are calculated to the maximum order $n$ as in \Eq{cc1_finite}, 
the $[p/q]$ Pad\'e approximant can be given as 
\beq
\girpq = c_1 \Delta_{N_f}\left[\frac{1+\sum_{j=1}^p b_j \Delta_{N_f}^j}{1 + \sum_{k=1}^q d_k \Delta_{N_f}^k}\right], 
\label{eq:pade_gir}
\eeq
with $p+q+1=n$. 
The coefficients are uniquely determined by matching the terms with those in \Eq{cc1_finite} 
after expanding $\girpq$ around $\Delta_{N_f}=0$. 
The differences among $[p/q]$ Pad\'e approximants are at $\mathcal{O}(\Delta_{N_f}^{(n+1)})$. 
With $n=4$, there are four possibilities: $[3/0]$, $[2/1]$, $[1/2]$, and $[0/3]$ Pad\'e approximants, 
where $[3/0]$ is nothing but the original conformal expansion $\gir^{(n=4)}(\Delta_{N_f})$. 

The $[p/q]$ Pad\'e approximant in \Eq{pade_gir} is a meromorphic function by construction, 
which is defined in a certain domain of the complex plane including the origin, with $q$ poles. 
To make it useful to our analysis, 
the Pad\'e approximant should not have a pole in $N_f^{\rm cr} \leq N_f \leq N_f^{\rm AF}$ 
such that it is well defined in the whole region of the conformal window.\footnote{
We note that the Pad\'e approximant for $\gir$ will still miss the expected squre-root singularity at $N_f^{\rm cr}$ by construction. 
In contrast, we expect that the Pad\'e approximant for $\Gir$ well approximates the closed form if exists. 
} 
In Ref.~\cite{Ryttov:2017lkz} it was suggested that the Pad\'e approximant must have no poles 
within the bound estimated by which $2$-loop beta function loses an IR fixed point. 
There, several exemplified $SU(N)$ theories coupled to $N_f$ fermions in the representation $R$ 
for some small integer values of $N$ and the large $N$ limit were extensively investigated, 
where the suggested condition has been used to find the valid Pad\'e approximants. 
However, the $2$-loop analysis is too naive to estimate $N_f^{\rm cr}$ and 
thus this criteria might be insufficient to find the valid Pad\'e approximants.

In this work, we determine the best Pad\'e approximant by 
exploiting the large-$N_f$ technique \cite{Gracey:1996he,Holdom:2010qs} as follows. 
For illustration purposes we consider the Veneziano limit of $SU(N)$ gauge theories 
coupled to $N_f$ fundamental fermion matter as in \Sec{critical_condition}. 
We first find all the four possible Pad\'e approximants defined in \Eq{pade_gir}. 
These approximants are supposed to extend the perturbative results beyond the perturbative regime near $x_f^{\rm AF}=11/2$. 
Note that for $x_f > x_f^{\rm AF}$ the perturbative BZ fixed points are found in the nonunitary regime with $\alpha_{\rm BZ}<0$. 
In the mean time, we compute the anomalous dimension $\gir(\lambda_{f,\,{\rm IR}})$ 
using the leading-order result of the large-$N_f$ expansion in the $\overline{\rm MS}$ scheme \cite{Espriu:1982pb,PalanquesMestre:1983zy}, 
where the (large-$N_f$) IR coupling $\lambda_{f,\,{\rm IR}}$ is determined from the two lowest terms of the large $N_f$ beta function 
\cite{Gracey:1996he,Holdom:2010qs}, 
i.e., $\beta(\lambda_{f})\simeq \frac{4 T_F}{3}\lambda_{f}^2+\frac{\beta^{(1)}(\lambda_f)}{N_f}=0$ 
with $\lambda_f=N_f \alpha/4\pi$.\footnote{
We refer the reader to the Appendix {\bf B} of Ref.~\cite{DiPietro:2020jne} for comprehensive results of $\gpsi$ and $\beta$ 
at $\mathcal{O}(N_f^{-1})$ in the large $N_f$ expansion.
} 
In the Veneziano limit, the coupling $\lambda_f$ is replaced by $x_f \alpha/4\pi$, 
where the large $N_f$ means the large $x_f$ limit. 
Similar to the coupling expansion, 
the large-$N_f$ analysis also finds a fixed point at the negative value of $\alpha$ for $x_f > x_f^{\rm AF}$.\footnote{
This fixed point should be distinguished from the UV fixed point at the positive value of $\alpha$, 
yielded by the singularity in $\beta^{(1)}(\lambda_f)$ at $\lambda_f=3/2$, in the context of asymptotic safety.\cite{Litim:2014uca}} 
If the Pad\'e approximant well approximates the anomalous dimension at the fixed point, 
we then expect that it should be in agreement with the large-$N_f$ result for a sufficiently large value of $x_f$. 
To be specific, as suggested in Ref.~\cite{Holdom:2010qs}, the large-$N_f$ expansion is supposed to be valid for $x_f \gtrsim 10$ 
provided that $|x_f\alpha/4\pi|\sim 3/2$. 
For $x_f > x_f^{\rm AF}$ we have a fixed point at the negative value of the large-$N_f$ coupling, 
where its magnitude increases from $0$ to $3/2$ as $x_f$ monotonically varies from $x_f^{\rm AF}=11/2$ to $\sim 15$. 
Beyond this we find an oscillatory behavior of the large-$N_f$ expansion, which is hardly captured by the Pad\'e approximants. 

We therefore find a limited window in $x_f$ , $10 \lesssim x_f \lesssim 15$, 
for which the large-$N_f$ expansion is under control 
and the resulting values of the anomalous dimension at the fixed point can be compared to the Pad\'e approximants. 
In the right panel of \Fig{dxf}, we show the results of $\gamma_{\rm BZ}$: 
the black solid line is for the large $N_f$ expansion, while 
the red, blue and green dashed lines are for $[3/0]$, $[1/2]$, and $[0/3]$ Pad\'e approximants, respectively. 
As seen in the figure, $[0/3]$ Pad\'e approximant is in good agreement with the large $N_f$ result 
over the aforementioned range. 
Although we do not present the results here, this qualitative picture holds 
for all the other finite and infinite $N$ theories considered in the next section. 
Therefore, we use $[0/3]$ Pad\'e approximant for our best estimate of $\girpq$ 
throughout this work.\footnote{
In most theories considered in this work $[0/3]$ Pad\'e approximant also satisfied the condition suggested in Ref.~\cite{Ryttov:2017lkz}. 
There are a few cases, however, that it has a pole slightly within the conformal boundary 
estimated by the $2$-loop beta function analysis. 
Nevertheless, we believe that our conclusion does not conflict with the restriction of which 
the Pad\'e approximant must not have a pole in the conformal window, 
because in general the would-be conformal window is expected to be narrower than the one determined 
from that the $2$-loop IR coupling runs to infinity. 
}

We carry out the same analysis for $\Gir$ and find that $[0/3]$ Pad\'e approximants also 
provide the best results as they qualitatively agree with the large $N_f$ results 
over the range in $x_f$ mentioned before. 
Furthermore, as shown in the left panel in \Fig{dxf}, 
it turns out that 
$[0/3]$ Pad\'e approximants obtained from both critical conditions 
result in better agreement than those at finite order. 
We therefore determine the critical number of flavors $N_f^{{\rm cr}_1,[0/3]}$ 
and $N_f^{{\rm cr}_2,[0/3]}$ from both critical conditions , 
$\gamma_{{\rm IR},[0/3]}(\Delta_{N_f})=1$ and $\left[\gir(2-\gir)\right]_{[0/3]}(\Delta_{N_f})=1$, 
respectively, and take the difference to $N_f^{\rm cr}$ as 
our estimate for the uncertainty 
related to the limitation of the finite-order critical condition:
\beq
\delta_2^{+} = N_f^{{\rm cr}_2,\,[0/3]} - N_f^{\rm cr}~~{\rm and}~~
\delta_2^{-} = N_f^{{\rm cr}_1,\,[0/3]} - N_f^{\rm cr},
\label{eq:delta_2}
\eeq
where we recall that $N_f^{\rm cr}$ is obtained from the critical condition defined in \Eq{cc2_finite} with $n=4$.

\section{Conformal window in nonsupersymmetric gauge theories}
\label{sec:cw}

In this section, we present our main results for the conformal window 
in nonsupersymmetric gauge theories coupled to fermionic matter 
using the strategies discussed in the previous section. 
In particular, we consider non-abelian gauge theories with $SU(N)$, $Sp(2N)$, and $SO(N)$ gauge groups 
and fermionic matter fields in the fundamental (F), adjoint (Adj), two-index symmetric (S2) 
and antisymmetric (AS) representations. 
In the case of $SO(N)$,  we also consider the spinorial (S) representation. 
The group invariants necessary for the computation of $k_{1,\,\cdots,\,4}$ in the conformal expansion of $\gir$ 
in \Eq{cc1_finite} are basically the same with the ones used for calculations of 
$\beta(\alpha)$ and $\gamma(\alpha)$ at the $4$th order in the coupling expansion \cite{vanRitbergen:1997va,Chetyrkin:1997dh,Vermaseren:1997fq}. 
The results have further been generalized to two-index and spinorial representations 
using the general fourth-order Casimir invariants for simple Lie groups \cite{Okubo:1981td,Okubo:1982dt}, 
where the detailed discussions and notations are found in the Appendices of Refs.~\cite{Pica:2010xq,Ryttov:2017dhd} 
and \cite{Mojaza:2010cm}, respectively. 
(See also Tables in the Appendix A of Ref.~\cite{Kim:2020yvr} for the summary of the resulting expressions 
relevant to this work.) 

The lower end of the conformal window has been estimated by a number of different analytical methods. 
Among those known in the literature, we consider the following three scheme-independent 
analytical approaches for a comparison:
\begin{itemize}
\item The $2$-loop beta function loses the BZ fixed point when $\alpha_{\rm BZ}=-4\pi b_1/b_2$ 
runs to infinity, which yields that
\beq
N_f^{{\rm cr,2-loop}}=\frac{17 C_2(G)^2}{T(R)\left[10 C_2(G)+6 C_2(R)\right]}.
\label{eq:ncr_2loop}
\eeq
\item The SD analysis in the ladder approximation suggests that the loss of the conformality 
happens if the IR coupling satisfies the condition, $\alpha_{\rm IR}=\alpha^{\rm cr}$ 
with $\alpha^{\rm cr}=\pi/3C_2(R)$. 
Conventionally the IR coupling is approximated by $\alpha_{\rm BZ}$ and one finds
\beq
N_f^{{\rm cr, SD}}=\frac{C_2(G)(17 C_2(G)+66 C_2(R))}{T(R) (10 C_2(G)+30 C_2(R))}.
\label{eq:ncr_sd}
\eeq
\item A closed form of the beta function was proposed in Ref.~\cite{Ryttov:2007cx}, 
which is scheme-independent in the sense that it involves 
the first two universal coefficients of $\beta(\alpha)$, 
the first universal coefficient of $\gpsi(\alpha)$, 
and a physical input for $\girpsi$.
\footnote{%
In Ref.~\cite{Pica:2010mt} a modified version of the all-order beta function was also proposed. 
By setting $\girpsi=1$, even though we do not present the results in this paper, 
we find that the resulting values of $N_f^{\rm cr}$ are similar to $N_f^{\rm cr, 2-loop}$ 
for the fundamental and spinorial representations, 
while they are roughly lying in the middle between $N_f^{\rm cr, 2-loop}$ and our results 
for other representations. 
}
To be consistent with the critical condition used for this work, we set $\girpsi=1$. 
It leads to a simple expression for the critical number of flavors
\beq
N_f^{\rm cr, BF}=\frac{11}{6}\frac{C_2(G)}{T(R)}.
\label{eq:ncr_bf}
\eeq
\end{itemize}
For some theories at finite $N$ we also compare our results with recent lattice results. 

\subsection{Large $N$ limit}
\label{sec:large_n}

Before we present the results at finite $N$, let us first consider an appropriate large $N$ limit 
in which the fermionic flavors are still relevant to the dynamics. 
As discussed in \Sec{conformal_phase}, 
we note that the critical condition to the anomalous dimension of the fermion bilinear in \Eq{critical_condition} 
would be exact at infinite $N$. 
For the fundamental flavors we take the Veneziano limit, 
i.e., both $N$ and $N_f$ are infinite while keeping $x_f=N_f/N$ finite. 
For $SU(N)$ and $Sp(2N)$ we find 
\beq
[x_f^{\rm cr},\,x_f^{\rm AF}]=\left[3.39^{+0.10}_{-0.09}\,^{+0.08}_{-0.14},\,5.5\right].
\label{eq:cw_fund_veneziano}
\eeq 
Throughout this work we denote the errors to $x_f^{\rm cr}$ and $N_f^{\rm cr}$ 
by $\,^{\delta_1^+}_{\delta_1^-}\,^{\delta_2^+}_{\delta_2^-}$, 
where $\delta_1^\pm$ and $\delta_2^\pm$ are defined in Eqs. \ref{eq:delta_1} and \ref{eq:delta_2}, respectively. 
For $SO(N)$ the conformal window is basically the same as that of $SU(N)$ 
except that now the Veneziano limit is defined with Weyl fermions, $x_{Wf}=N_{Wf}/N$, 
\beq
[x_{Wf}^{\rm cr},x_{Wf}^{\rm AF}]=\left[3.39^{+0.10}_{-0.09}\,^{+0.08}_{-0.14},5.5\right].
\eeq 
For a comparison 
we also present the resulting values of $x_f^{\rm cr}$ ($x_{Wf}^{\rm cr}$) defined in 
Eqs.~\ref{eq:ncr_2loop}, \ref{eq:ncr_sd}, and \ref{eq:ncr_bf}
\beq
x_f^{\rm cr, 2-loop} = \frac{34}{13},~x_f^{\rm cr, SD} = 4,~{\rm and} ~x_f^{\rm cr, BF} = \frac{11}{3}.
\eeq

In the cases of the adjoint and two-index symmetric and antisymmetric representations, 
we take the 't Hooft large $N$ limit, i.e. $N\rightarrow \infty$ with fixed $N_f$. 
For the adjoint representation the results are same for all the three gauge groups, 
which is also true for the other analytical methods mentioned above, and we find
\beq
[N_f^{\rm cr},\,N_f^{\rm AF}]=\left[1.90^{+0.03}_{-0.03}\,^{<0.01}_{-0.11},\,2.75\right], 
\label{eq:cw_su_adj_largeN}
\eeq
or equivalently, 
\beq
[N_{Wf}^{\rm cr},\,N_{Wf}^{\rm AF}]=\left[3.79^{+0.07}_{-0.07}\,^{+0.01}_{-0.23},\,5.5\right].
\label{eq:cw_adj_largeN}
\eeq
Using Eqs.~\ref{eq:ncr_2loop}, \ref{eq:ncr_sd}, and \ref{eq:ncr_bf}, 
we also find that
\beq
N_f^{\rm cr, 2-loop} = \frac{17}{16},~N_f^{\rm cr, SD} = \frac{83}{40}, ~{\rm and}~N_f^{\rm cr, BF} = \frac{11}{6}.
\label{eq:cw_anal_adj}
\eeq

For both two-index symmetric and antisymmetric representations of $SU(N)$ the conformal windows are 
exactly twice larger than that for the adjoint representation 
\beq
[N_f^{\rm cr},\,N_f^{\rm AF}]=\left[3.79^{+0.07}_{-0.07}\,^{+0.01}_{-0.23},\,5.5\right].
\label{eq:cw_two_index_largeN}
\eeq
In the case of $Sp(2N)$ with fermions in the antisymmetric representation, the conformal window 
is equivalent to that for the adjoint representation. 
Similarly, for $SO(N)$ the conformal window for the symmetric representation is same with that for the adjoint representation. 
Note that in these theories the other two-index representations are identical to the adjoint representation by construction. 
Analogously, the other analytical calculations for the two-index representations 
yield the same results of \Eq{cw_anal_adj} up to a factor of two.

\subsection{$SU(N)$ gauge theory with $N_f$ Dirac fermions in various representations}
\label{sec:cw_su}

\begin{figure}
\begin{center}
\includegraphics[width=0.45\textwidth]{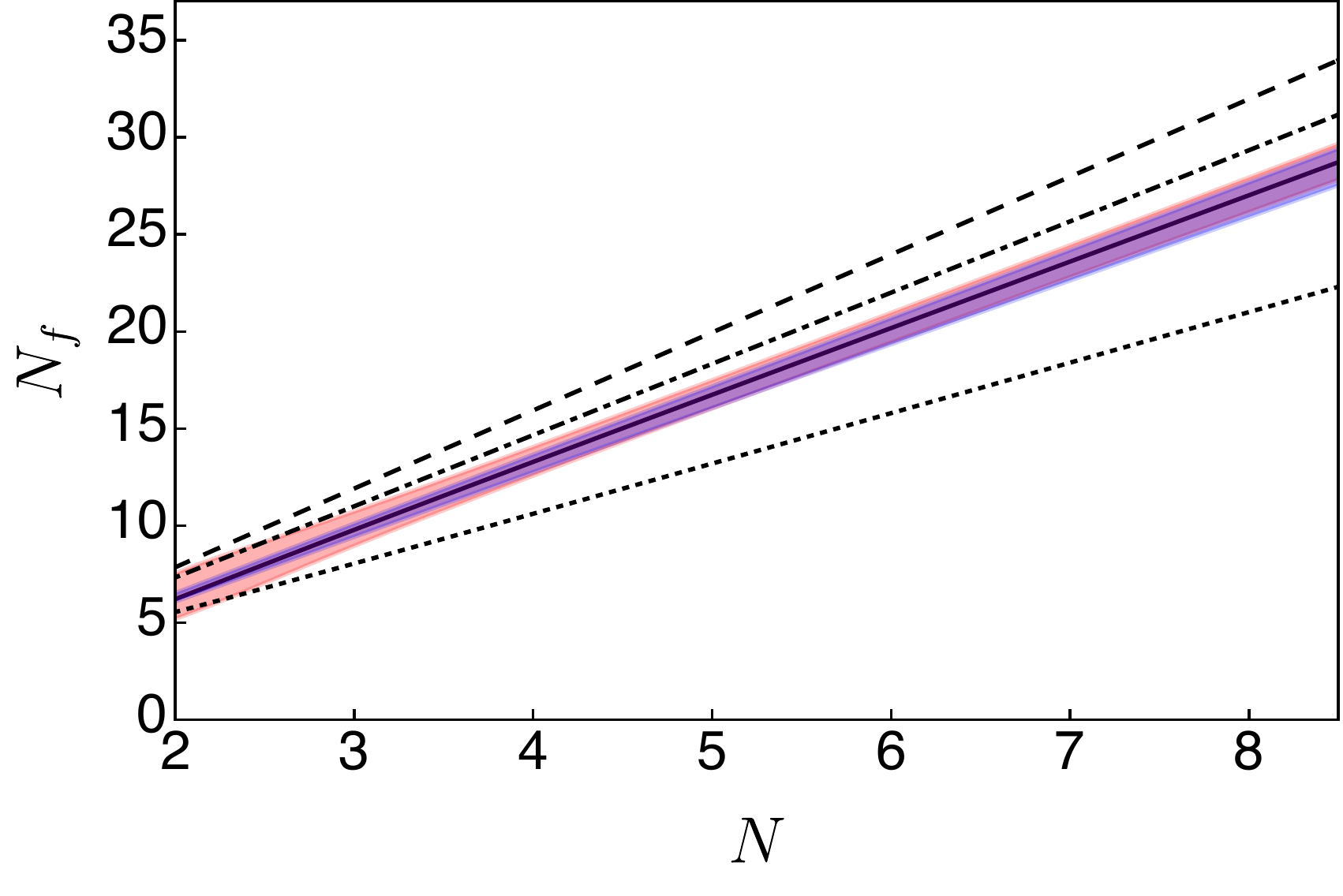}
\includegraphics[width=0.45\textwidth]{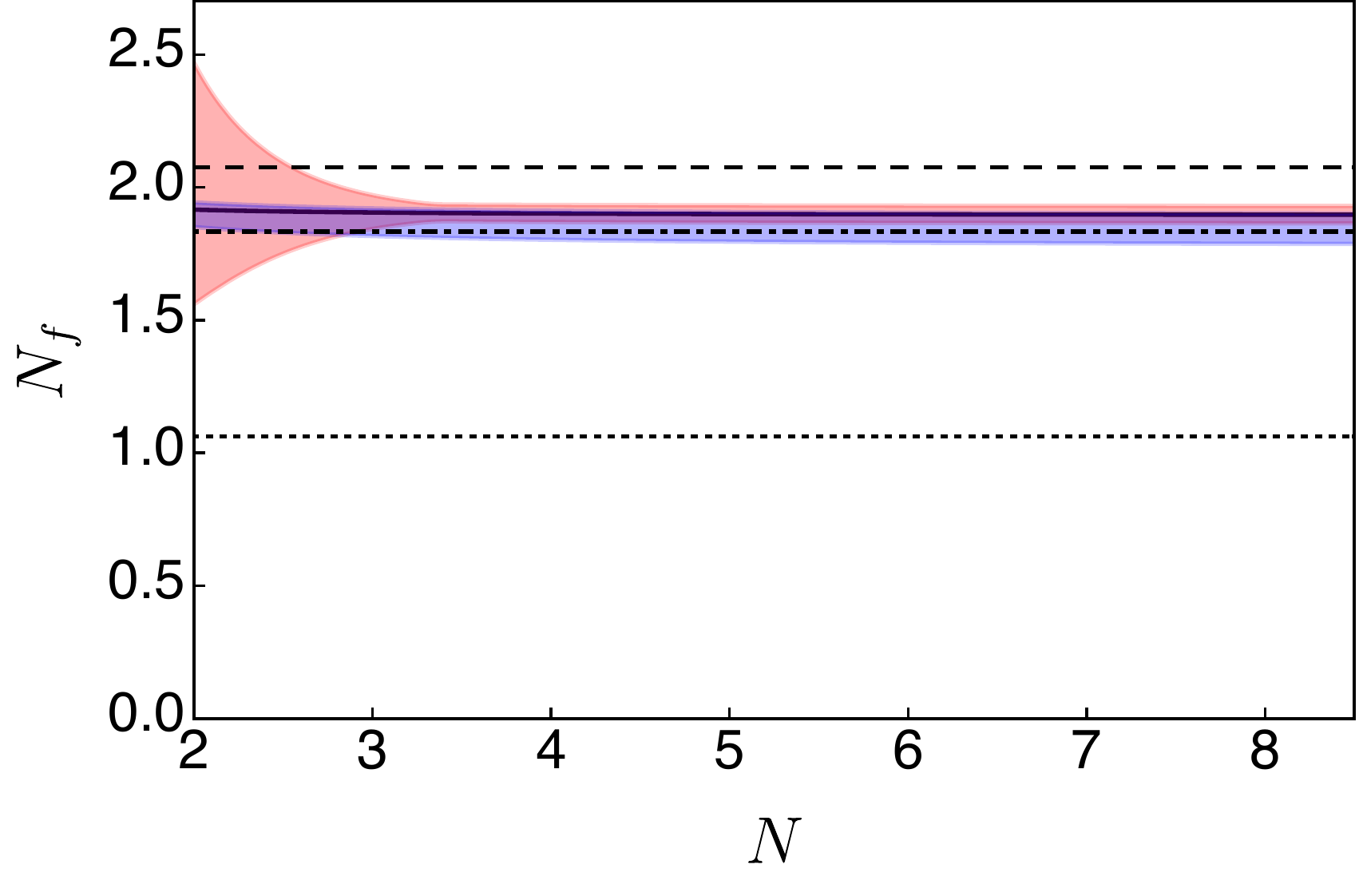}
\includegraphics[width=0.45\textwidth]{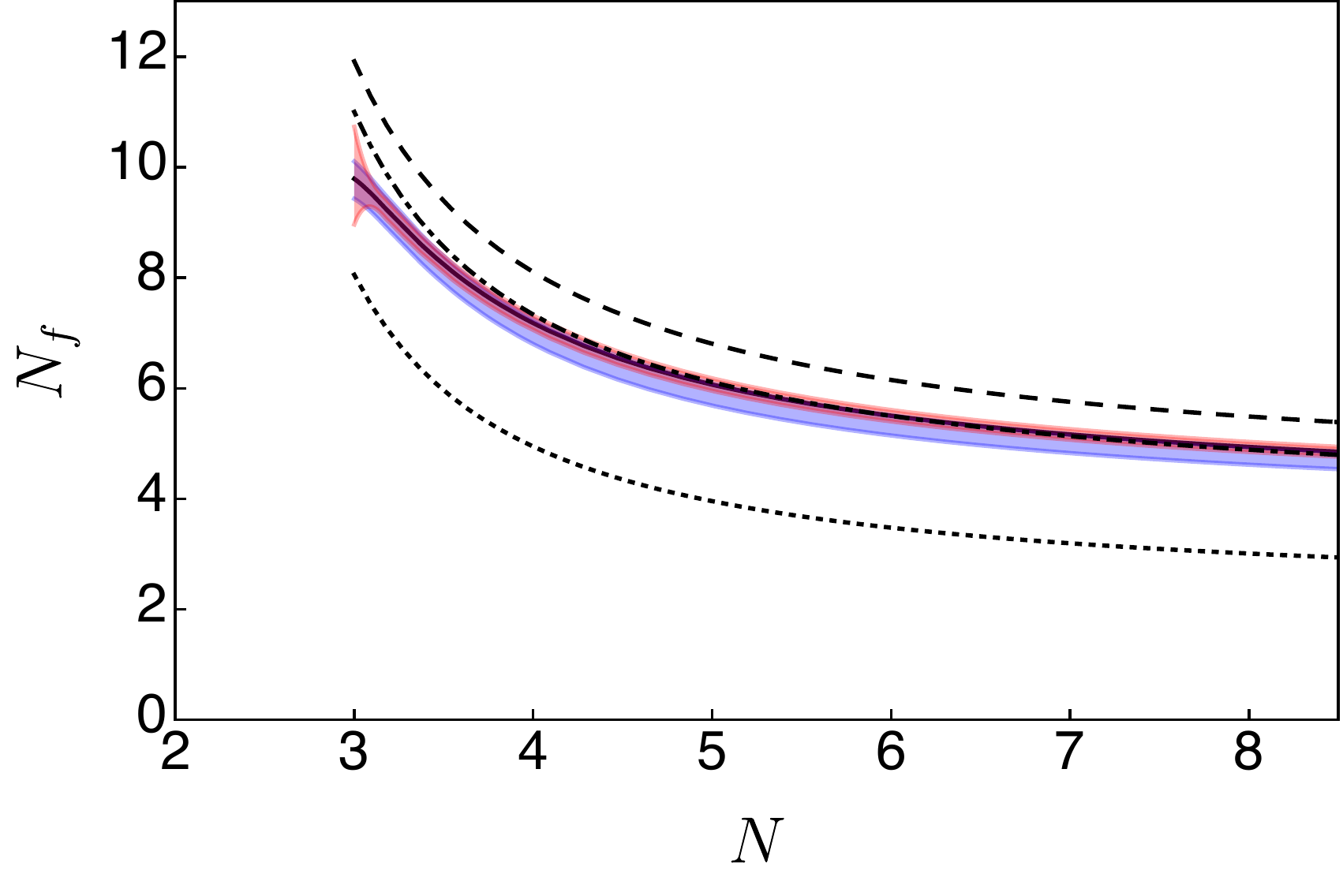}
\includegraphics[width=0.45\textwidth]{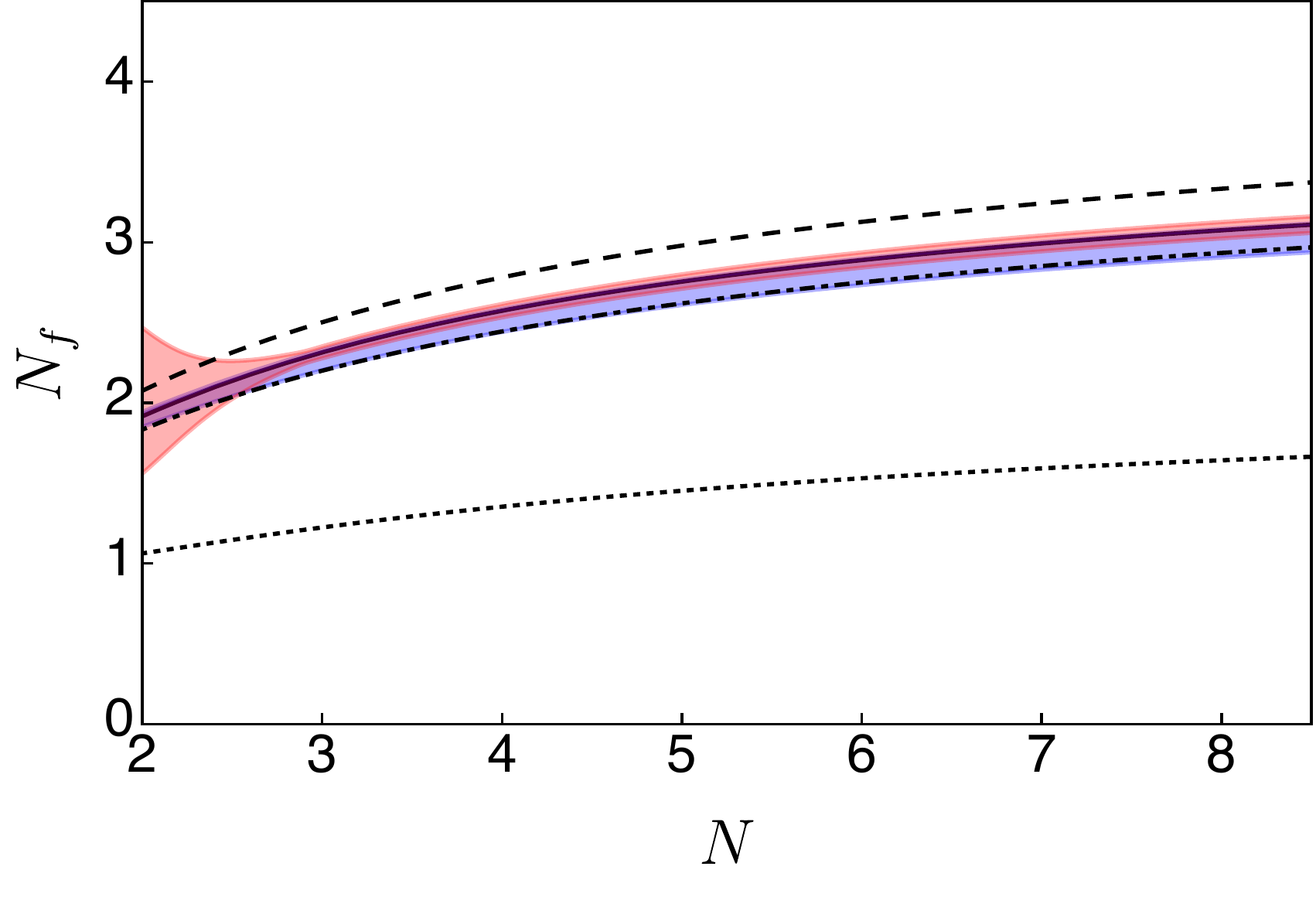}
\caption{%
The boundary between conformal and chirally broken phases in $SU(N)$ gauge theories 
with $N_f$ Dirac flavors of fermion in the fundamental (top-left), 
adjoint (top-right), two-index antisymmetric (bottom-left) and symmetric (bottom-right) representations. 
The black solid line is estimated from the finite-order critical condition 
in \Eq{cc2_finite} with $n=4$, 
where red and blue bands denote the systematic errors computed according to Eqs.~\ref{eq:delta_1} and \ref{eq:delta_2}, respectively. 
Dotted, dashed, and dot-dashed lines are for the analytical results estimated by the $2$-loop beta function, 
the truncated Schwinger-Dyson, and the all-order beta function analyses 
in Eqs.~\ref{eq:ncr_2loop}, \ref{eq:ncr_sd}, and \ref{eq:ncr_bf}. 
}
\label{fig:su_cw_reps}
\end{center}
\end{figure}

\begin{table}
\small
\begin{center}
\begin{tabular}{|c|c|c|c|c|}
\hline\hline
$N$ & F & Adj & AS & S2 \\
\hline
2 & $\left[6.22\,^{+1.32}_{-1.01}\,^{+0.31}_{-0.12},\,11.0\right]$ & $\left[1.92\,^{+0.55}_{-0.36}\,^{+0.03}_{-0.07},\,2.75\right]$ 
& N/A & $\left[1.92\,^{+0.55}_{-0.36}\,^{+0.03}_{-0.07},\,2.75\right]$\\
3 & $\left[9.79\,^{+0.94}_{-0.82}\,^{+0.31}_{-0.36},\,16.5\right]$ & $\left[1.91\,^{+0.07}_{-0.06}\,^{+0.02}_{-0.09},\,2.75\right]$ 
& $\left[9.79\,^{+0.94}_{-0.82}\,^{+0.31}_{-0.36},\,16.5\right]$ & $\left[2.31\,^{+0.04}_{-0.04}\,^{+0.01}_{-0.11},\,3.3\right]$\\
4 & $\left[13.29\,^{+0.77}_{-0.71}\,^{+0.37}_{-0.53},\,22.0\right]$ & $\left[1.90\,^{+0.03}_{-0.03}\,^{+0.01}_{-0.10},\,2.75\right]$ 
& $\left[7.18\,^{+0.13}_{-0.13}\,^{+0.07}_{-0.38},\,11.0\right]$ & $\left[2.57\,^{+0.04}_{-0.04}\,^{+0.01}_{-0.13},\,3.67\right]$\\
5 & $\left[16.74\,^{+0.75}_{-0.70}\,^{+0.44}_{-0.68},\,27.5\right]$ & $\left[1.90\,^{+0.03}_{-0.03}\,^{+0.01}_{-0.10},\,2.75\right]$ 
& $\left[\,6.06^{+0.11}_{-0.11}\,^{+0.02}_{-0.38},\,9.17\right]$ & $\left[2.75^{+0.05}_{-0.04}\,^{+0.01}_{-0.15},\,3.93\right]$\\
6 & $\left[20.18\,^{+0.79}_{-0.74}\,^{+0.51}_{-0.83},\,33.0\right]$ & $\left[1.90\,^{+0.03}_{-0.03}\,^{+0.01}_{-0.11},\,2.75\right]$ 
& $\left[5.50^{+0.10}_{-0.10}\,^{+0.01}_{-0.36},\,8.25\right]$ & $\left[2.89\,^{+0.05}_{-0.05}\,^{+0.01}_{-0.16},\,4.13\right]$\\
7 & $\left[23.60\,^{+0.84}_{-0.80}\,^{+0.59}_{-0.97},\,38.5\right]$ & $\left[1.90\,^{+0.03}_{-0.03}\,^{+0.01}_{-0.11},\,2.75\right]$ 
& $\left[5.16^{+0.10}_{-0.09}\,^{+0.01}_{-0.34},\,7.7\right]$ & $\left[2.99\,^{+0.05}_{-0.05}\,^{+0.01}_{-0.16},\,4.28\right]$\\
8 & $\left[27.01\,^{+0.91}_{-0.87}\,^{+0.66}_{-1.12},\,44.0\right]$ & $\left[1.90\,^{+0.03}_{-0.03}\,^{<0.01}_{-0.11},\,2.75\right]$ 
& $\left[4.94^{+0.09}_{-0.09}\,^{+0.01}_{-0.32},\,7.33\right]$ & $\left[3.07\,^{+0.05}_{-0.05}\,^{+0.01}_{-0.17},\,4.4\right]$\\
9 & $\left[30.42\,^{+0.99}_{-0.94}\,^{+0.74}_{-1.26},\,49.5\right]$ & $\left[1.90\,^{+0.03}_{-0.03}\,^{<0.01}_{-0.11},\,2.75\right]$ 
& $\left[4.78^{+0.09}_{-0.09}\,^{<0.01}_{-0.31},\,7.07\right]$ & $\left[3.14\,^{+0.05}_{-0.05}\,^{+0.01}_{-0.17},\,4.5\right]$\\
10 & $\left[33.83\,^{+1.07}_{-1.02}\,^{+0.82}_{-1.40},\,55.0\right]$ & $\left[1.90\,^{+0.01}_{-0.03}\,^{<0.01}_{-0.11},\,2.75\right]$ 
& $\left[4.65^{+0.09}_{-0.08}\,^{<0.01}_{-0.30},\,8.88\right]$ & $\left[3.17\,^{+0.05}_{-0.05}\,^{+0.01}_{-0.18},\,4.58\right]$\\
\hline\hline
\end{tabular}
\end{center}
\caption{
\label{tab:su_cw}%
Conformal window of $SU(N)$ gauge theories coupled to fermion matter 
in the fundamental (F), adjoint (Adj), antisymmetric (AS), and symmetric (S2) representations. 
The lower and upper bounds, denoted by [$N_f^{\rm cr}$, $N_f^{\rm AF}$], correspond to which 
conformality and asymptotic freedom are lost, respectively.  
The first and second errors to $N_f^{\rm cr}$ are computed according to Eqs.~\ref{eq:delta_1} and \ref{eq:delta_2}, respectively.
}
\end{table}

In \Fig{su_cw_reps} we present our results for the lower edge of the conformal window in $SU(N)$ gauge theories 
coupled to $N_f$ Dirac fermions in the fundamental, adjoint, antisymmetric and symmetric representations, 
where the resulting values of $N_f^{\rm cr}$ are denoted by black solid lines. 
Note that we take both $N$ and $N_f$ as continuous variables. 
In each figure, red and blue bands 
denote the errors, $\delta_1$ and $\delta_2$, defined in Eqs.~\ref{eq:delta_1} and \ref{eq:delta_2}, respectively. 
We recall that these errors should not be taken simultaneously 
since the underlying assumptions for the error estimates are incompatible to each other as discussed in \Sec{error_estimate}. 
For the integer values of $N$ ranged over $2 \leq N \leq 10$ we present the explicit values of $N_f^{\rm cr}$ 
with errors in \Tab{su_cw}, where the values of $N_f^{\rm AF}$ are also presented. 

As shown in the figures, $\delta_2$ persists to be sizable for all the values of $N$ at $\sim6\%$ level at most. 
On the other hand, $\delta_1$ is relatively large at small $N$, but comparable or smaller than $\delta_2$ at large $N$. 
If we concern the truncation error $\delta_1$, 
such a result implies that $SU(N)$ theories at small $N$ receive significant nonperturbative corrections 
near the lower end of the conformal window, 
which result in the large ambiguity of the perturbative conformal series expansion. 

We compare our results to other analytical methods in the figures: dotted lines are for the $2$-loop beta function analysis, 
dashed lines for the traditional SD method, and dot-dashed lines for the all-order beta function with $\gir=1$. 
We find that our results are in between $N_f^{\rm cr, 2-loop}$ and $N_f^{\rm cr, SD}$, 
and more or less comparable to $N_f^{\rm cr, BF}$, except the fundamental representation at large $N$, 
if the errors $\delta_2$ are concerned. 
Note that the adjoint and symmetric representations are identical for $N=2$, while 
the antisymmetric and fundamental representations are identical for $N=3$. 

Studying the infrared dynamics of an interacting (near) conformal theory 
has also been a rich subject of lattice gauge theories, 
because the corresponding IR coupling is typically in the strong coupling regime and 
thus it requires reliable nonperturbative techniques. 
However, it is a highly nontrivial task to investigate the deep IR regime of the {\it would-be} conformal theories 
using lattice simulations due to the large finite-size effects 
as expected from the fact that the correlation length diverges at an IR fixed point. 
Nevertheless, in recent years various numerical techniques have been developed to tackle such a problem, 
and turned out to be successful as they revealed various aspects of the (near) conformal dynamics from first principles. 
(See Sec.~V in \cite{Brower:2019oor} for a brief summary 
of the recent developments and challenges along this direction.)

In particular, $SU(3)$ theories with many fundamental flavors have been extensively studied, 
because they are not only easily utilizing the state-of-the-art lattice techniques 
justified by successfully simulating QCD, but also provide useful benchmark studies 
for more interesting UV models in the context of physics beyond the standard model. 
Although it is not yet conclusive, 
the most recent lattice results 
suggest that $N_f=12$ is conformal \cite{Hasenfratz:2019dpr},\footnote{
See also, e.g., \cite{Fodor:2017gtj} for a different point of view. 
} 
$8$ is chirally broken but nearly conformal \cite{Appelquist:2018yqe}, 
and $10$ is likely conformal but controversial \cite{Chiu:2018edw,Fodor:2018tdg,Hasenfratz:2020ess,Appelquist:2020xua}. 
In the case of $SU(2)$, 
much evidence has been found that $N_f=4$ is chirally broken while $6$ is likely conformal, 
e.g., see \cite{Leino:2018yfd} and references therein. 
As shown in \Tab{su_cw}, our results are in excellent agreement with these lattice results for both cases.\footnote{
For $SU(3)$ more analytical results, besides the ones used for a comparison in \Sec{cw}, are available in the literature. 
In particular, our findings are consistent with the results in Refs.~\cite{Gies:2005as,Antipin:2018asc}, 
but different with those in Refs.~\cite{Kusafuka:2011fd,Kuipers:2018lux}. 
} 

\begin{figure}
\begin{center}
\includegraphics[width=0.59\textwidth]{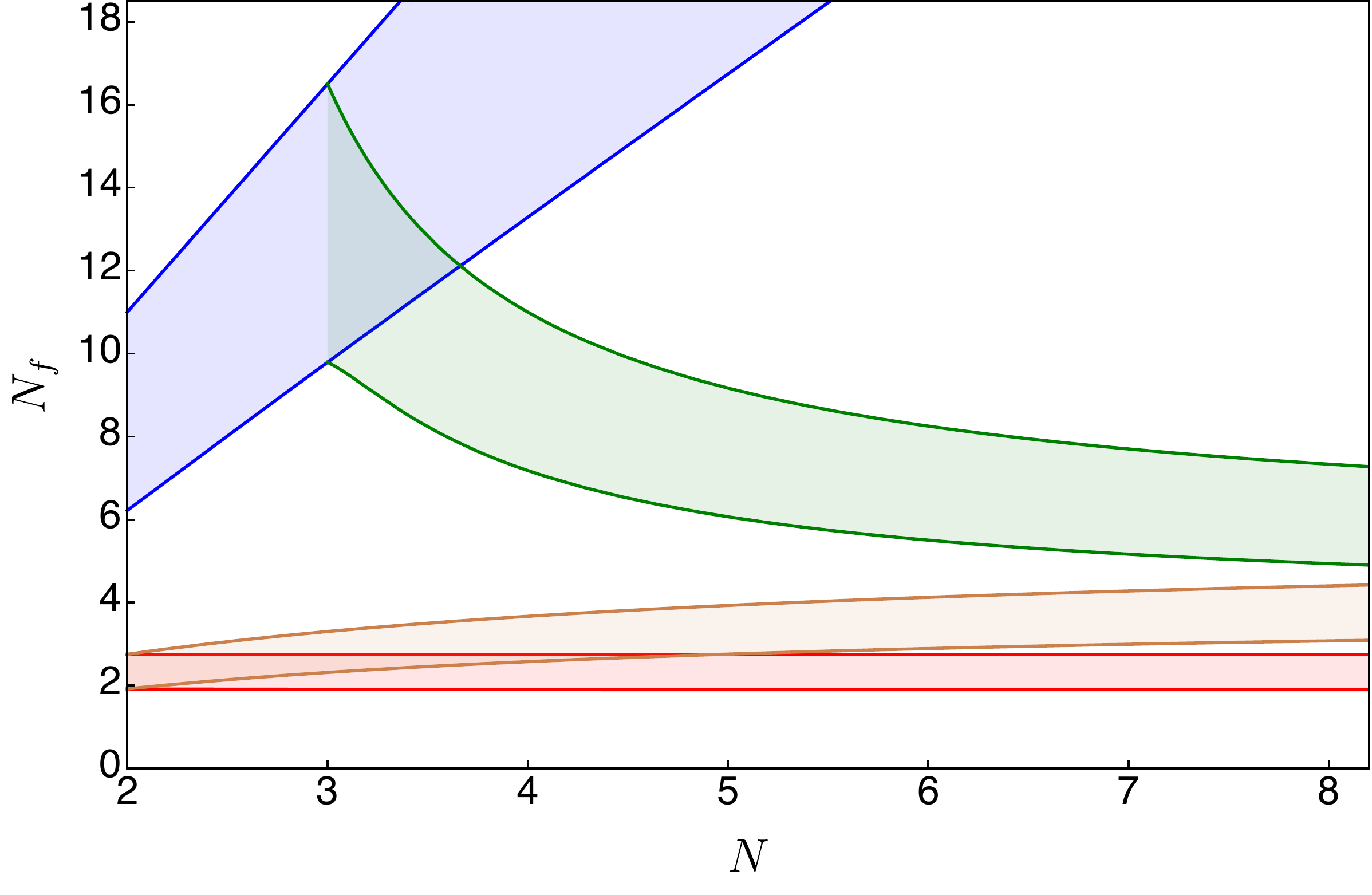}
\caption{%
Conformal window of $SU(N)$ gauge theory with $N_f$ Dirac flavors of fermion in the fundamental (blue), 
adjoint (red), antisymmetric (green), and symmetric (brown) representations. 
The upper bound is determined by the perturbative beta function in the onset of the loss of an asymptotic freedom, 
while the lower bound is estimated from the finite-order critical condition 
truncated at the $4$th order of the conformal expansion defined in \Eq{cc2_finite}. 
}
\label{fig:su_cw}
\end{center}
\end{figure}

Besides the fundamental representation, adjoint and two-index representations have also been studied 
by the means of nonperturbative lattice calculations. 
Lattice studies of $SU(2)$ with adjoint fermion matter find evidence for that $N_f=2,\,3/2$ are conformal, 
$1/2$ is chirally broken, and $1$ is likely conformal (e.g., see \cite{Bergner:2019kub} and references therein), 
which is somewhat different to what we have found. 
However, we note that the adjoint $SU(2)$ seems to receive significant nonperturbative corrections, 
as indicated by the large truncation errors shown in the top-right panel of \Fig{su_cw_reps} and \Tab{su_cw}. 
As we discussed in \Sec{conformal_phase}, 
it is also possible that for $N=2$ 
the critical condition in \Eq{critical_condition} could largely be modified 
by receiving a substantial amount of finite $N$ corrections. 
Furthermore, recently it was suggested that $SU(2)$ theories with $N_f=1$ and $3/2$ adjoint fermions 
might exhibit an exotic phase in the deep IR, confining but no chiral symmetry breaking \cite{Anber:2018iof,Poppitz:2019fnp}. 
Hence, nothing is conclusive for these specific theories yet and it would be interesting to further investigate 
the nonperturbative effects in detail by improving the error analyses. 
$SU(3)$ theory with $2$ symmetric (sextet) Dirac fermions has also been studied extensively by lattice techniques, 
as it has been shown to exhibit near conformal behaviors like as in the $8$ fundamental-flavor $SU(3)$ theories 
(for the most recent progress see Ref.~\cite{Fodor:2020niv} and references therein). 
Our results strongly support such lattice results. 

We summarize our findings on the conformal window of $SU(N)$ gauge theories 
with $N_f$ flavors of fermion in the fundamental, adjoint, two-index symmetric and antisymmetric representations in \Fig{su_cw}.

\subsection{$Sp(2N)$ gauge theory with $N_f$ Dirac fermions in various representations}
\label{sec:cw_sp}

\begin{figure}
\begin{center}
\includegraphics[width=0.45\textwidth]{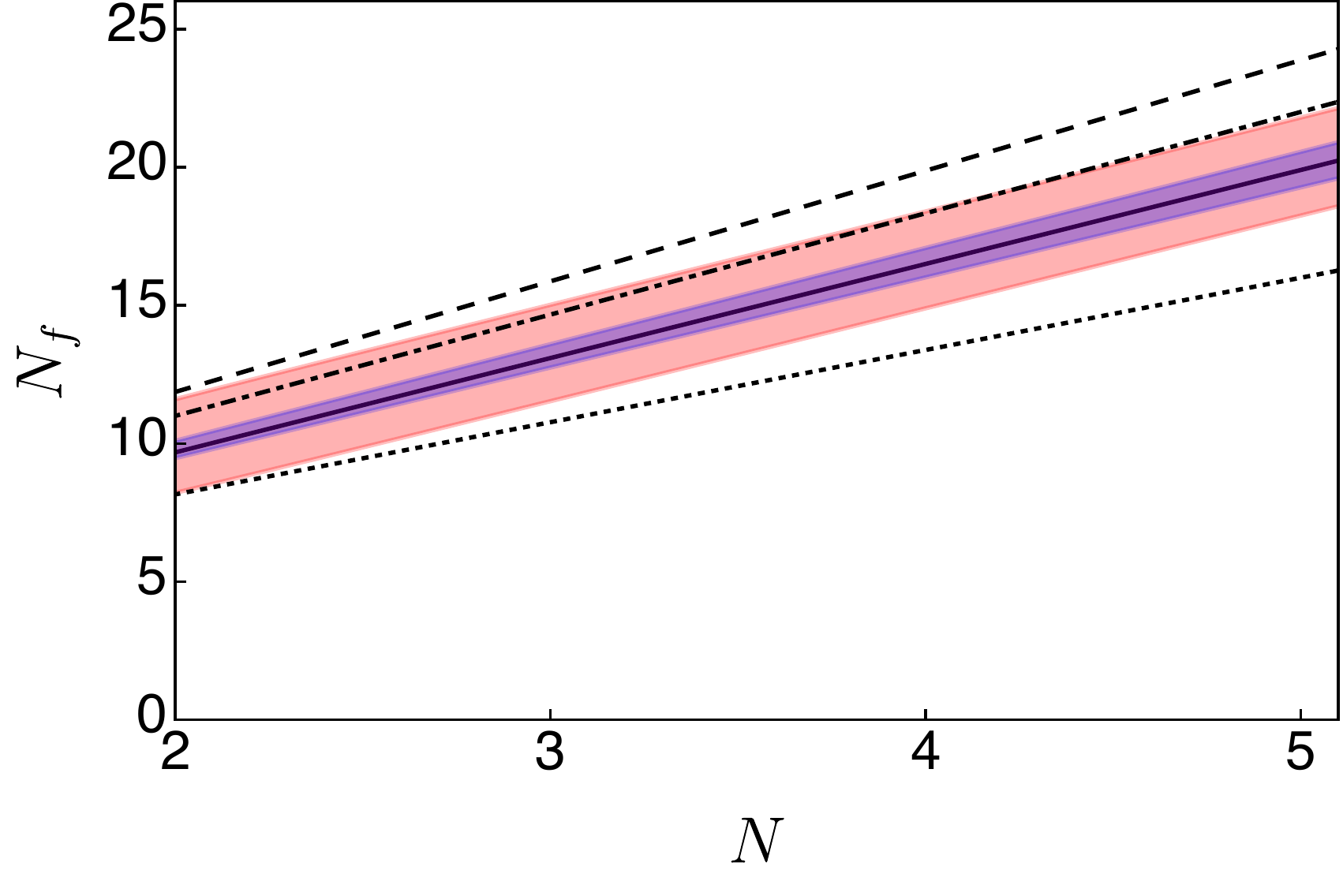}
\includegraphics[width=0.45\textwidth]{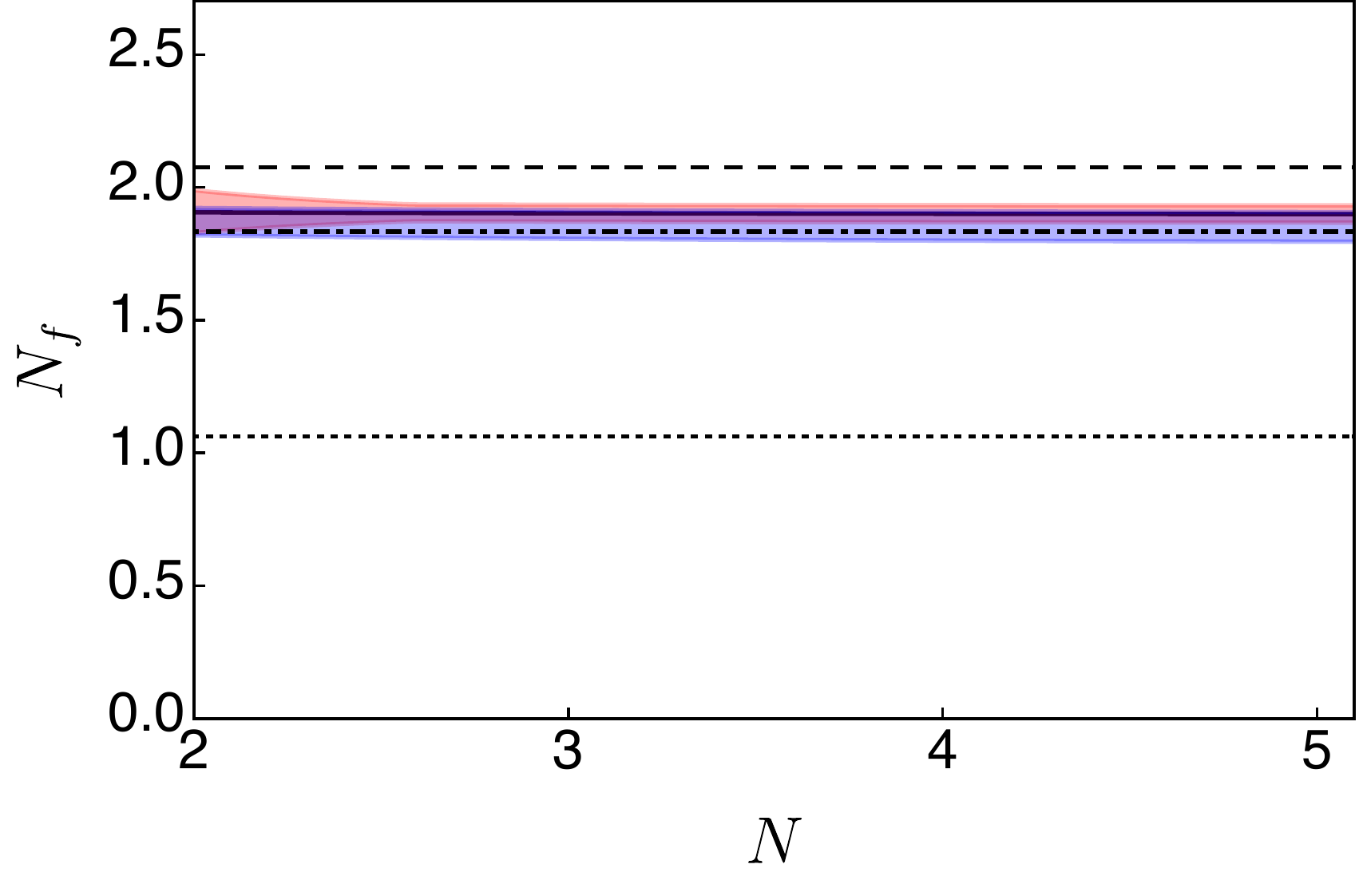}
\includegraphics[width=0.45\textwidth]{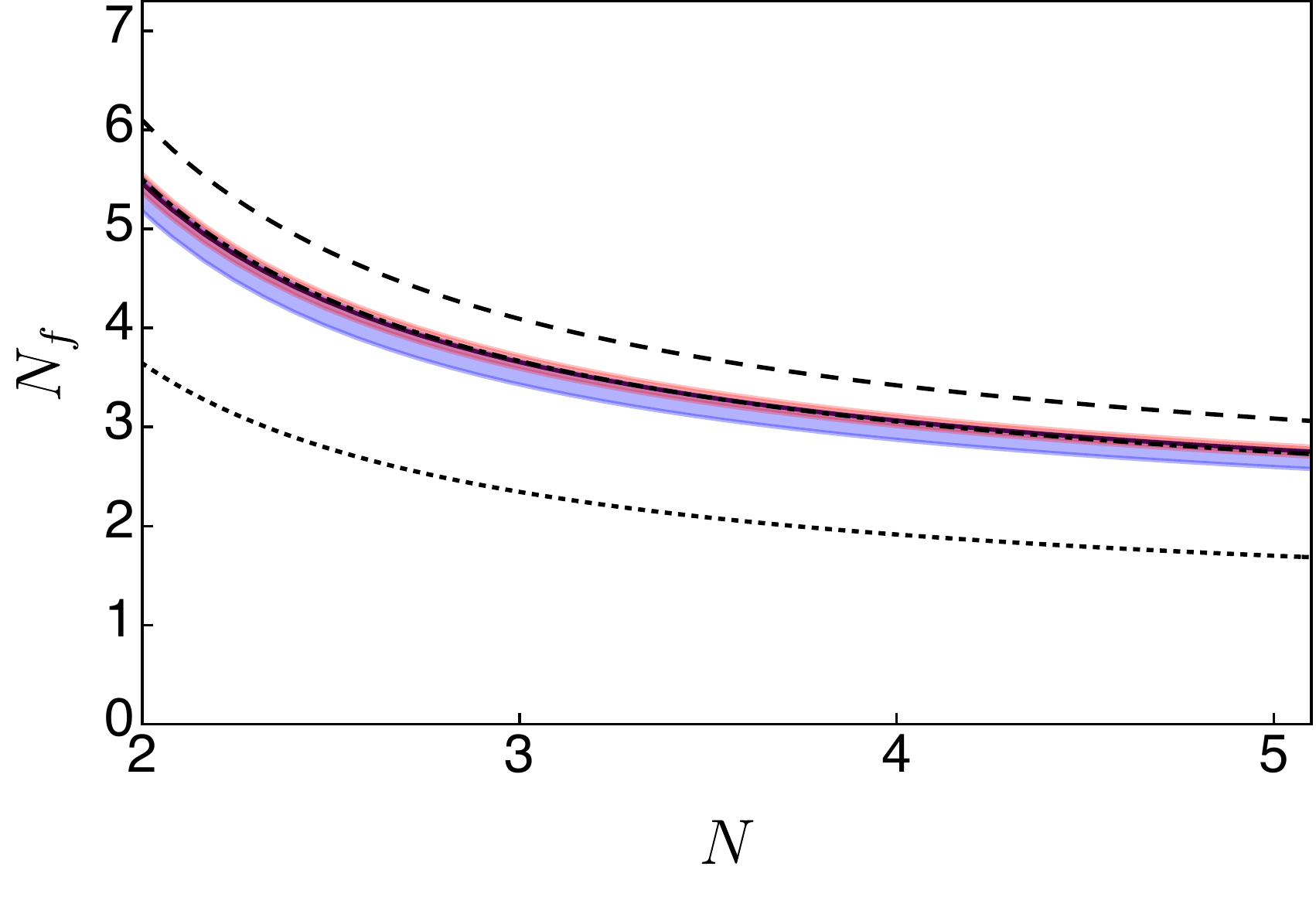}
\caption{%
The boundary between conformal and chirally broken phases in $Sp(2N)$ gauge theories 
with $N_f$ Dirac flavors of fermion in the fundamental (top-left), 
adjoint (top-right) and two-index antisymmetric (bottom) representations. 
The black solid line is estimated from the finite-order critical condition 
in \Eq{cc2_finite} with $n=4$, 
where red and blue bands denote the systematic errors computed according to Eqs.~\ref{eq:delta_1} and \ref{eq:delta_2}, respectively. 
Dotted, dashed, and dot-dashed lines are for the analytical results estimated by the $2$-loop beta function, 
the truncated Schwinger-Dyson, and the all-order beta function analyses 
in Eqs.~\ref{eq:ncr_2loop}, \ref{eq:ncr_sd}, and \ref{eq:ncr_bf}. 
}
\label{fig:sp_cw_reps}
\end{center}
\end{figure}

In the case of $Sp(2N)$ we consider the fundamental, 
adjoint ($=$ two-index symmetric) and two-index antisymmetric representations. 
We present our results with black solid lines in \Fig{sp_cw_reps}, 
where the red and blue bands denote the errors, 
$\delta_1$ and $\delta_2$, defined in Eqs.~\ref{eq:delta_1} and \ref{eq:delta_2}, respectively. 
Again, we take $N$ and $N_f$ as continuous variables to obtain the results in the figure. 
For several small integer values of $N=2,\,\cdots,\,6$, we present the resulting values of $N_f^{\rm cr}$ with errors 
and $N_f^{\rm AF}$ in \Tab{sp_cw}.
Note that $Sp(2)=SU(2)$. 
For a comparison, we also present the results of other analytical approaches 
defined in Eqs.~\ref{eq:ncr_2loop}, \ref{eq:ncr_sd} and \ref{eq:ncr_bf}. 
The generic trend is similar to what we found for $SU(N)$, except that the truncation errors $\delta_1$ 
are much larger than $\delta_2$ and those of $SU(N)$ in the fundamental representation at the same values of $N$. 

Nonperturbative lattice studies of $Sp(2N)$ gauge groups are barely found in the literature. 
Only recently has a research program for $Sp(4)$ lattice theories with fermions in the fundamental and antisymmetric representations 
begun by aiming to explore the composite dynamics of the electroweak symmetry breaking and composite dark matter \cite{Bennett:2017kga,Bennett:2019jzz,Bennett:2019cxd}. 
Our results suggest that $Sp(4)$ theory would exhibit near conformal behavior 
if it couples to $8\sim 9$ fundamental or $4\sim 5$ antisymmetric flavors of fermion.\footnote{%
A phenomenologically interesting minimal $Sp(4)$ composite Higgs model requires to contain $2$ fundamental and $3$ antisymmetric fermions 
\cite{Barnard:2013zea}. 
In this case, analytical studies using the same strategy used for this work, but truncated at the $3$rd order 
in the conformal expansion of $\gir$ for multiple representations, suggest that the model 
resides slightly outside the conformal window \cite{Kim:2020yvr}.
} 
Since the lattice studies of $Sp(2N)$ at large $N$ are also being pursued by the same research group \cite{Holligan:2019lma}, 
we should note that $Sp(6)$ and $Sp(8)$ with $3$ antisymmetric fermions 
would be good candidates for near conformal theories.  

\begin{table}
\small
\begin{center}
\begin{tabular}{|c|c|c|c|}
\hline\hline
$N$ & F & Adj & AS  \\
\hline
2 & $\left[9.68\,^{+1.92}_{-1.46}\,^{+0.42}_{-0.21},\,16.5\right]$ & $\left[1.91\,^{+0.08}_{-0.08}\,^{+0.02}_{-0.09},\,2.75\right]$ 
& $\left[5.46\,^{+0.10}_{-0.09}\,^{+0.05}_{-0.28},\,8.25\right]$\\
3 & $\left[13.09\,^{+1.93}_{-1.56}\,^{+0.50}_{-0.34},\,22.0\right]$ & $\left[1.90\,^{+0.03}_{-0.03}\,^{+0.01}_{-0.10},\,2.75\right]$ 
& $\left[3.66\,^{+0.07}_{-0.07}\,^{+0.01}_{-0.23},\,5.5\right]$\\
4 & $\left[16.50\,^{+1.90}_{-1.61}\,^{+0.58}_{-0.49},\,27.5\right]$ & $\left[1.90\,^{+0.03}_{-0.03}\,^{+0.01}_{-0.10},\,2.75\right]$ 
& $\left[3.07\,^{+0.06}_{-0.06}\,^{+0.01}_{-0.19},\,4.58\right]$\\
5 & $\left[19.90\,^{+1.90}_{-1.65}\,^{+0.66}_{-0.63},\,33.0\right]$ & $\left[1.90\,^{+0.03}_{-0.03}\,^{+0.01}_{-0.10},\,2.75\right]$ 
& $\left[2.77^{+0.05}_{-0.05}\,^{+0.01}_{-0.18},\,4.13\right]$\\
6 & $\left[23.30\,^{+1.92}_{-1.70}\,^{+0.74}_{-0.77},\,38.5\right]$ & $\left[1.90\,^{+0.03}_{-0.03}\,^{+0.01}_{-0.11},\,2.75\right]$ 
& $\left[2.60\,^{+0.05}_{-0.05}\,^{<0.01}_{-0.16},\,3.85\right]$\\
\hline\hline
\end{tabular}
\end{center}
\caption{
\label{tab:sp_cw}%
Conformal window of $Sp(2N)$ gauge theories with fermion matter 
in the fundamental (F), adjoint (Adj), and antisymmetric (AS) representations. 
The lower and upper bounds, denoted by [$N_f^{\rm cr}$, $N_f^{\rm AF}$], correspond to which 
conformality and asymptotic freedom are lost, respectively. 
The first and second errors to $N_f^{\rm cr}$ are computed according to Eqs.~\ref{eq:delta_1} and \ref{eq:delta_2}, respectively.
}
\end{table}

\begin{figure}
\begin{center}
\includegraphics[width=0.59\textwidth]{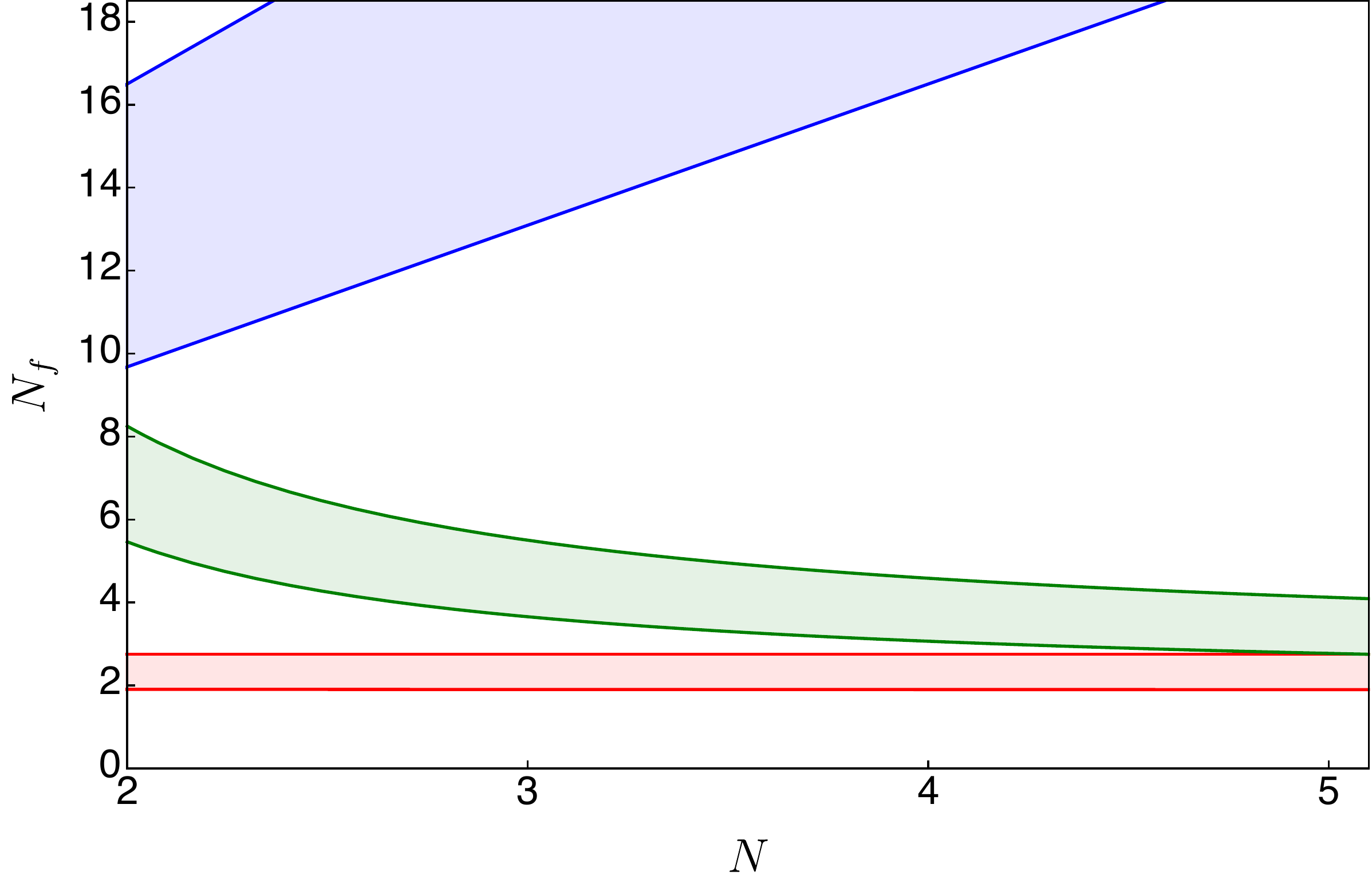}
\caption{%
Conformal window of $Sp(2N)$ gauge theory with $N_f$ Dirac flavors of fermion in the fundamental (blue), 
adjoint (red), and antisymmetric (green) representations. 
The upper bound is determined by the perturbative beta function in the onset of the loss of an asymptotic freedom, 
while the lower bound is estimated from the finite-order critical condition 
truncated at the $4$th order of the conformal expansion defined in \Eq{cc2_finite}. 
}
\label{fig:sp_cw}
\end{center}
\end{figure}

We summarize our findings on the conformal window of $Sp(2N)$ gauge theories 
coupled to $N_f$ flavors of fermion in the fundamental, adjoint and two-index antisymmetric representations in \Fig{sp_cw}. 

\subsection{$SO(N)$ gauge theory with $N_{Wf}$ Weyl fermions in various representations}
\label{sec:cw_so}

\begin{figure}
\begin{center}
\includegraphics[width=0.45\textwidth]{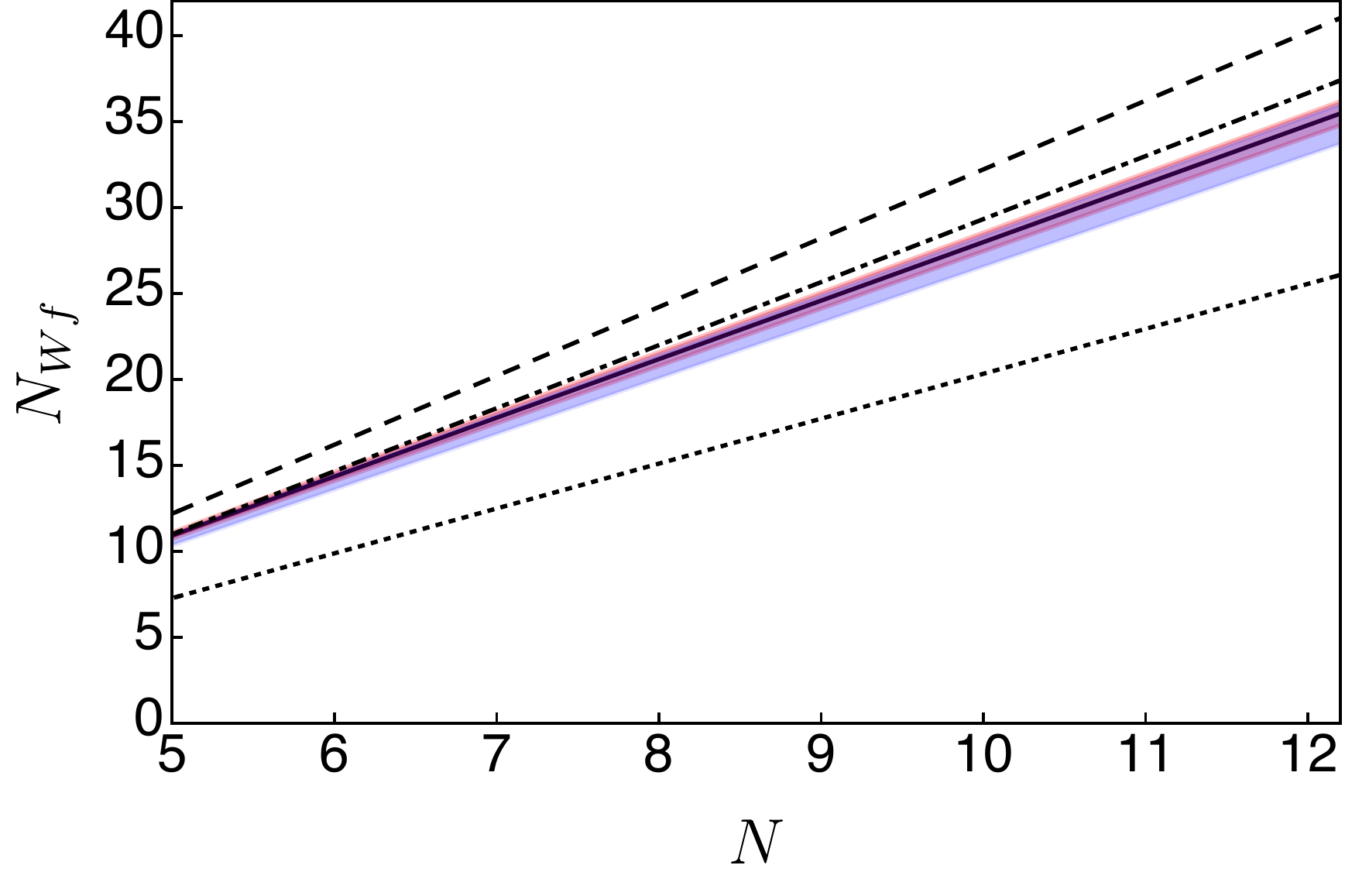}
\includegraphics[width=0.45\textwidth]{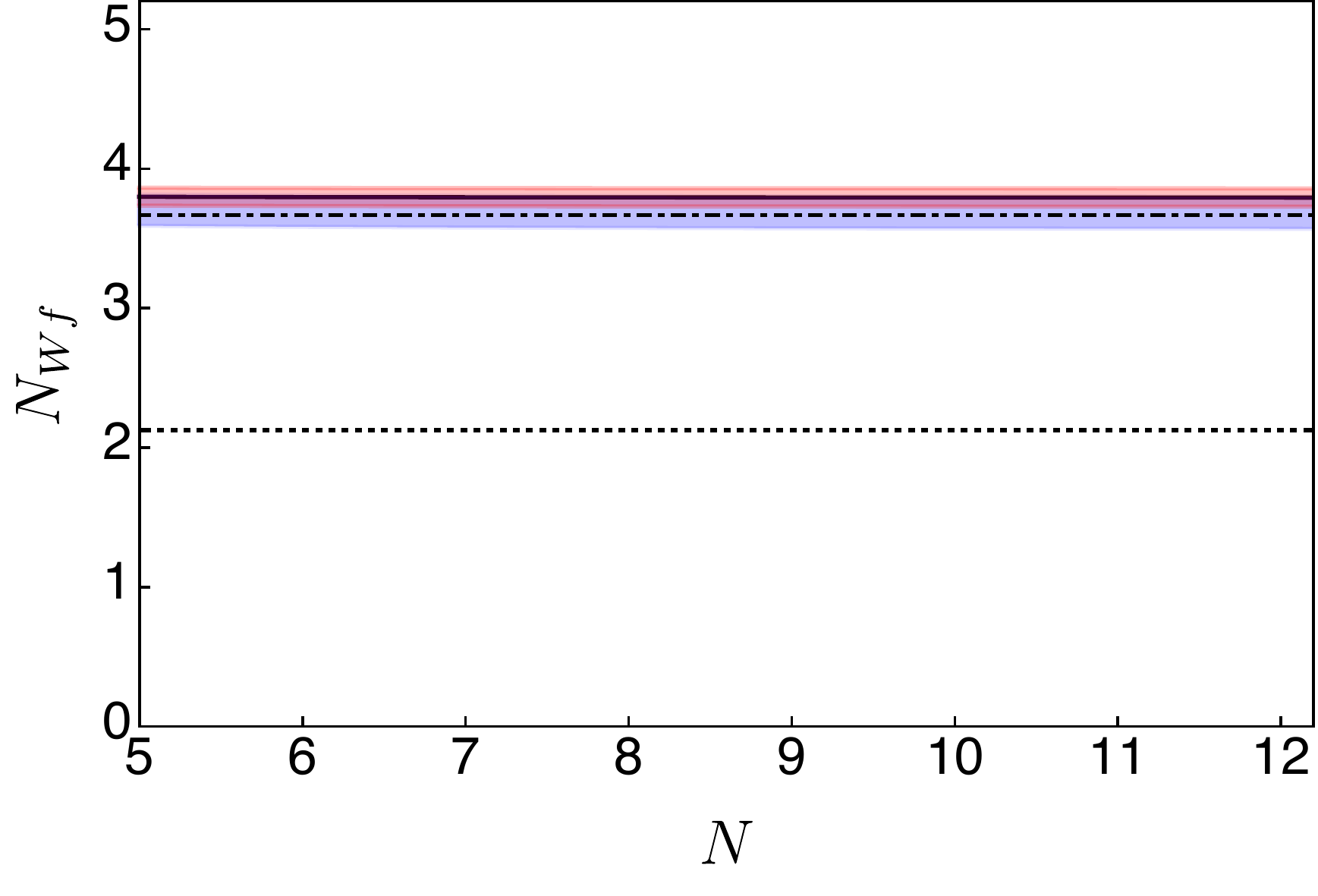}
\includegraphics[width=0.45\textwidth]{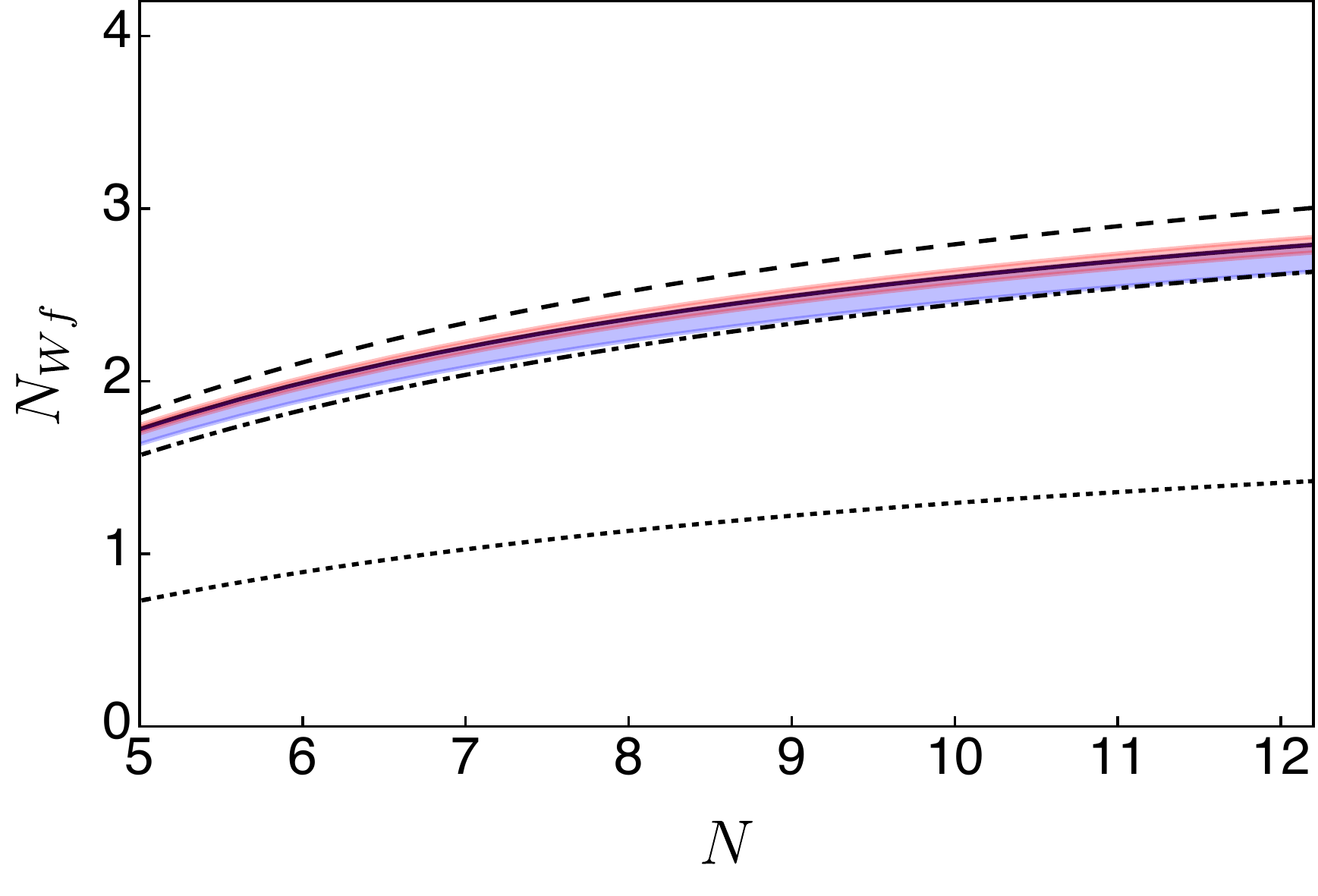}
\includegraphics[width=0.45\textwidth]{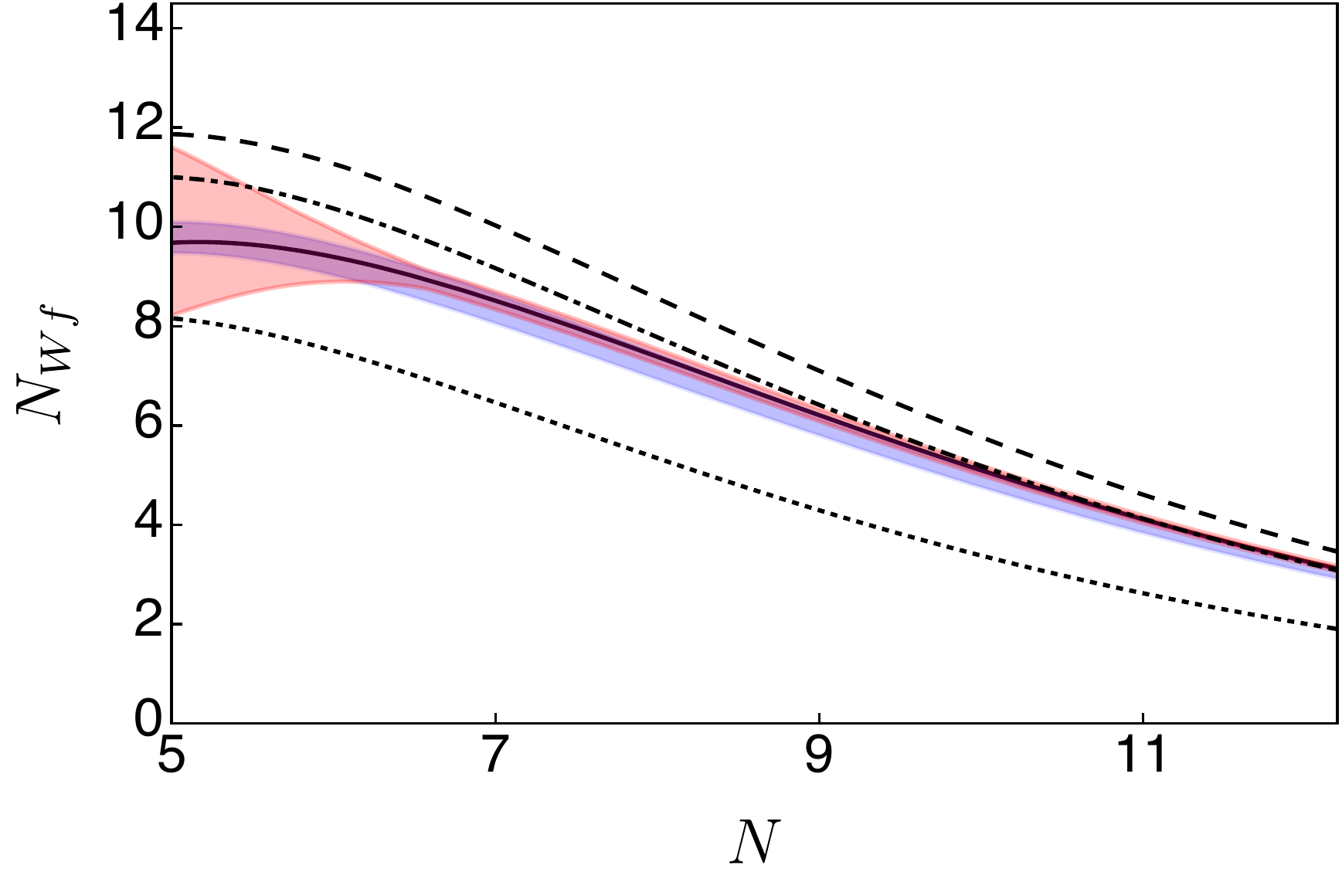}
\includegraphics[width=0.45\textwidth]{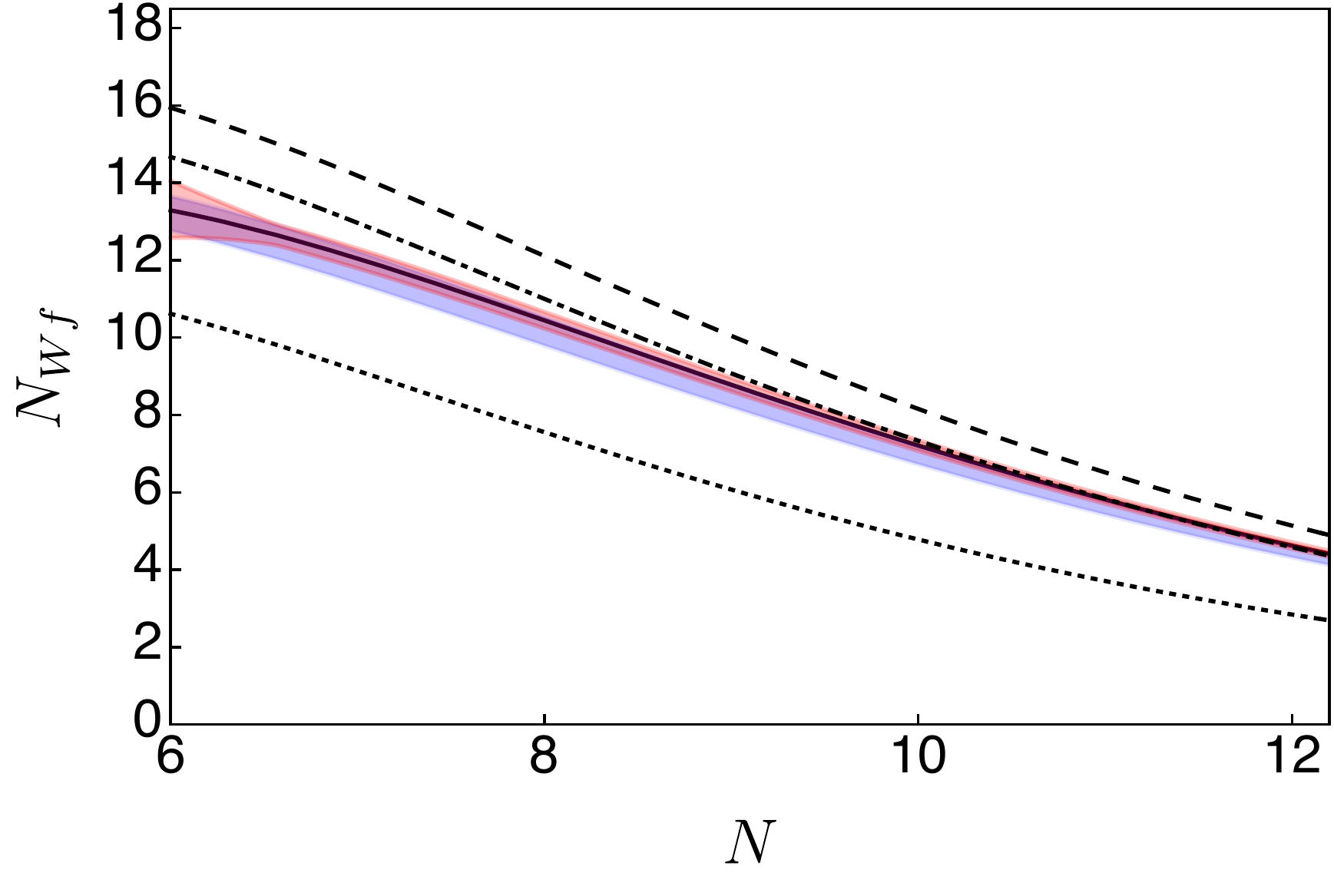}
\caption{%
The boundary between conformal and chirally broken phases in $SO(N)$ gauge theories 
with $N_{Wf}$ Weyl flavors of fermion in the fundamental (top-left), 
adjoint (top-right), two-index symmetric (middle-left) representations. 
In the case of the spinorial representation, we show the results for odd (middle-right) 
and even (bottom) integer values of $N$, separately. 
The black solid line is estimated from the finite-order critical condition 
in \Eq{cc2_finite} with $n=4$, 
where red and blue bands denote the systematic errors computed according to Eqs.~\ref{eq:delta_1} and \ref{eq:delta_2}, respectively. 
Dotted, dashed, and dot-dashed lines are for the analytical results estimated by the $2$-loop beta function, 
the truncated Schwinger-Dyson, and the all-order beta function analyses 
in Eqs.~\ref{eq:ncr_2loop}, \ref{eq:ncr_sd}, and \ref{eq:ncr_bf}. 
}
\label{fig:so_cw_reps}
\end{center}
\end{figure}

For $SO(N)$ gauge theories we consider $N_{Wf}$ Weyl fermions in the spinorial representation 
in addition to the fundamental, adjoint($=$ antisymmetric) and symmetric representations. 
We also restrict our attention to $N\geq 6$ since the results for the smaller values of $N$ 
could be deduced from $SU(2)$ and $Sp(4)$ gauge theories using the fact that $SO(3)\sim SU(2)$, $SO(4)\sim SU(2)\times SU(2)$ 
and $SO(5)\sim Sp(4)$ for which only even numbers of $N_{Wf}$ are allowed to avoid a Witten anomaly \cite{Witten:1982fp}. 

\begin{table}
\small
\begin{center}
\begin{tabular}{|c|c|c|c|c|}
\hline\hline
$N$ & F & Adj & S2 & S \\
\hline
6 & $\left[14.36\,^{+0.26}_{-0.26}\,^{+0.14}_{-0.76},\,22.0\right]$ & $\left[3.80\,^{+0.07}_{-0.06}\,^{+0.01}_{-0.21},\,5.5\right]$ 
& $\left[1.99\,^{+0.03}_{-0.03}\,^{<0.01}_{-0.10},\,2.75\right]$ & $\left[13.29\,^{+0.77}_{-0.71}\,^{+0.37}_{-0.53},\,22.0\right]$\\
7 & $\left[17.78\,^{+0.33}_{-0.33}\,^{+0.20}_{-0.94},27.5\right]$ & $\left[3.80\,^{+0.07}_{-0.06}\,^{+0.01}_{-0.22},\,5.5\right]$ 
& $\left[2.20\,^{+0.03}_{-0.03}\,^{<0.01}_{-0.11},\,3.06\right]$ & $\left[8.52\,^{+0.18}_{-0.18}\,^{+0.14}_{-0.46},\,13.75\right]$\\
8 & $\left[21.19\,^{+0.41}_{-0.40}\,^{+0.26}_{-1.11},\,33.0\right]$ & $\left[3.79\,^{+0.07}_{-0.06}\,^{+0.01}_{-0.22},\,5.5\right]$ 
& $\left[2.36^{+0.04}_{-0.04}\,^{<0.01}_{-0.12},\,3.3\right]$ & $\left[10.44^{+0.22}_{-0.21}\,^{+0.09}_{-0.65},\,16.5\right]$\\
9 & $\left[24.59\,^{+0.48}_{-0.47}\,^{+0.32}_{-1.27},\,38.5\right]$ & $\left[3.79\,^{+0.07}_{-0.06}\,^{+0.01}_{-0.22},\,5.5\right]$ 
& $\left[2.49^{+0.04}_{-0.04}\,^{<0.01}_{-0.13},\,3.5\right]$ & $\left[6.21\,^{+0.13}_{-0.12}\,^{+0.03}_{-0.41},\,9.63\right]$\\
10 & $\left[27.99\,^{+0.55}_{-0.54}\,^{+0.39}_{-1.43},\,44.0\right]$ & $\left[3.79\,^{+0.07}_{-0.06}\,^{+0.01}_{-0.22},\,5.5\right]$ 
& $\left[2.60^{+0.04}_{-0.04}\,^{<0.01}_{-0.14},\,3.67\right]$ & $\left[7.21\,^{+0.14}_{-0.14}\,^{+0.02}_{-0.48},\,11.0\right]$\\
11 & $\left[31.39\,^{+0.62}_{-0.60}\,^{+0.47}_{-1.58},\,49.5\right]$ & $\left[3.79\,^{+0.07}_{-0.06}\,^{+0.01}_{-0.22},\,5.5\right]$ 
& $\left[2.70^{+0.04}_{-0.04}\,^{<0.01}_{-0.15},\,3.81\right]$ & $\left[4.11\,^{+0.08}_{-0.08}\,^{+0.01}_{-0.27},\,6.19\right]$\\
12 & $\left[34.79\,^{+0.69}_{-0.67}\,^{+0.54}_{-1.74},\,55.0\right]$ & $\left[3.79\,^{+0.07}_{-0.06}\,^{+0.01}_{-0.22},\,5.5\right]$ 
& $\left[2.78^{+0.04}_{-0.04}\,^{<0.01}_{-0.15},\,3.93\right]$ & $\left[4.62\,^{+0.09}_{-0.08}\,^{<0.01}_{-0.30},\,6.88\right]$\\
\hline\hline
\end{tabular}
\end{center}
\caption{%
\label{tab:so_cw}%
Conformal window of $SO(N)$ gauge theories with fermion matter 
in the fundamental (F), adjoint (Adj), two-index symmetric (S2), and spinorial (S) representations. 
The lower and upper bounds, denoted by [$N_f^{\rm cr}$, $N_f^{\rm AF}$], correspond to which 
conformality and asymptotic freedom are lost, respectively. 
The first and second errors to $N_f^{\rm cr}$ are computed according to Eqs.~\ref{eq:delta_1} and \ref{eq:delta_2}, respectively.
}
\end{table}

We present our results for the lower edge of the conformal window by black solid lines in \Fig{so_cw_reps}. 
The blue and red bands denote the errors, $\delta_1$ and $\delta_2$, defined in Eqs.~\ref{eq:delta_1} and \ref{eq:delta_2}, respectively. 
Again, the results in the figures are obtained by treating $N$ and $N_f$ as continuous variables, 
but the physical system should take the integer values of $N$. 
In particular, the results for the spinorial representation displayed in the right-middle and the bottom panels of \Fig{so_cw_reps} 
are physical only at odd and even integer values of $N$, respectively. 

Our results are further compared to other analytical approaches, 
where the dotted lines are the results obtained from the loss of IR fixed point in $2$-loop beta function, 
the dashed lines from the traditional Schwinger-Dyson analysis, 
and the dot-dashed lines from the all-order beta function with $\gir=1$. 
Similar to $SU(N)$ and $Sp(2N)$, our results are comparable to $N_f^{\rm cr, BF}$, 
but smaller than $N_f^{\rm cr, SD}$ and larger than $N_f^{\rm cr, 2-loop}$, respectively. 
For several integer values of $N$ over the region, $6\leq N \leq 12$, we report our results in \Tab{so_cw}. 
Except in the case of the spinorial representation for $SO(6)$, $\delta_2$ is larger than $\delta_1$ 
for all the considered theories. 
For the adjoint representation the conformal windows are identical to those of $SU(N)$ gauge theories. 
In the case of $SO(6)$, the result of the spinorial representation is identical to that of the fundamental representation of $SU(4)$ 
up to a factor of $2$, as expected from $SO(6)\sim SU(4)$. 

\begin{figure}
\begin{center}
\includegraphics[width=0.59\textwidth]{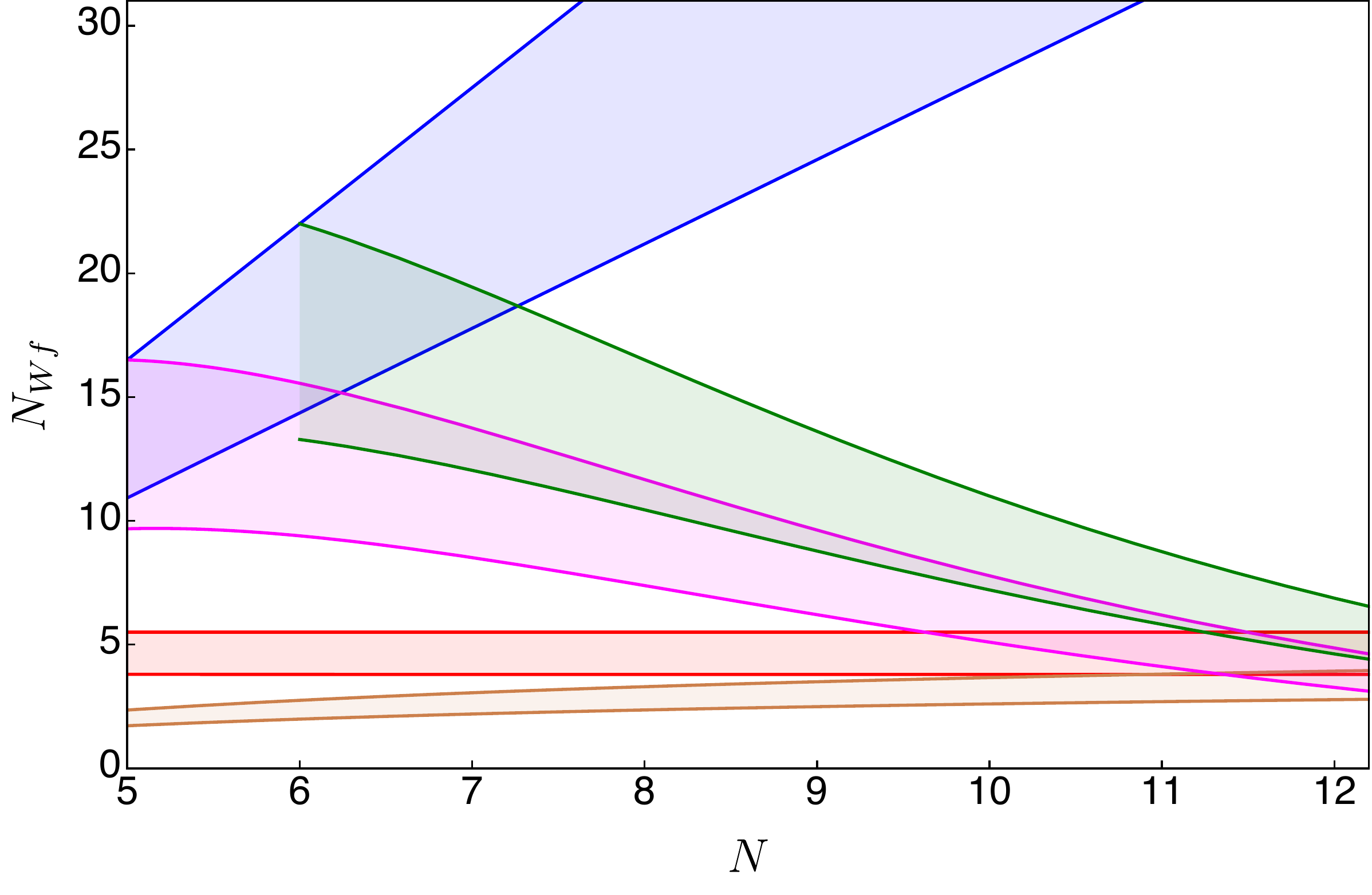}
\caption{%
Conformal window of $SO(N)$ gauge theory with $N_{Wf}$ Weyl flavors of fermion in the fundamental (blue), 
adjoint (red), symmetric (brown), 
spinorial (purple and green are for odd and even integer values of $N$, respectively) representations. 
The upper bound is determined by the perturbative beta function in the onset of the loss of an asymptotic freedom, 
while the lower bound is estimated from the finite-order critical condition 
truncated at the $4$th order of the conformal expansion defined in \Eq{cc2_finite}. 
}
\label{fig:so_cw}
\end{center}
\end{figure}

We summarize our findings on the conformal window of $SO(N)$ gauge theories 
coupled to $N_{Wf}$ flavors of fermion in the fundamental, adjoint, two-index symmetric and spinorial representations in \Fig{so_cw}.

\section{Conclusion}
\label{sec:conclusion}

We have investigated the conformal window of asymptotically free $SU(N)$, $Sp(2N)$ and $SO(N)$ gauge theories 
coupled to $N_f$ flavors of fermion in various representations, 
where their appropriate large $N$ limits are also considered. 
The upper end of the conformal window is identical to which asymptotic freedom is lost, 
and the corresponding number of flavors $N_f^{\rm AF}$ is determined by the perturbative RG beta function as usual. 
To find the lower end of the conformal window, 
in this work we have adopted the conjectured critical condition on the anomalous dimension of 
a fermion bilinear at an IR fixed point, $\girpsi$, 
in which the theory is expected to lose the conformality due to the dynamically generated scale (walking scaling) 
and fall into a chirally broken phase. 
This critical condition is expected to work well in the infinite $N$ limit, 
but could receive sizable corrections at finite $N$. 
We calculate $\girpsi$ by employing 
the Banks-Zaks conformal expansion 
whose expansion parameter $\Delta_{N_f}=N_f^{\rm AF}-N_f$, as well as the coefficient at each order, 
is free from the renormalization scheme.  
Following the strategy proposed in Ref.~\cite{Kim:2020yvr}, 
we have used the finite-order definition of the critical condition in \Eq{cc2_finite} 
up to the $4$th order, the highest order available up to date, 
to determine the critical number of flavors, $N_f^{\rm cr}$, 
which corresponds to the onset of the conformality lost. 
In particular, in Secs.~\ref{sec:conformal_phase} and \ref{sec:critical_condition} we discussed and emphasized 
why the latter definition of the critical condition in \Eq{critical_condition} works better than the former by providing both theoretical and empirical reasons. 

The highlight of this work lies in the estimation of the uncertainties 
in the resulting values of $N_f^{\rm cr}$ 
by treating the following two scenarios separately: the conformal series expansion is either convergent or divergent asymptotic. 
For the latter we have searched for the optimal truncation 
by assuming the factorial growth of the coefficients at larger order. 
Such divergent asymptotic behavior is related to the singularity in the Borel plane. 
We find that the truncation order used for the determination of $N_f^{\rm cr}$ 
was roughly the optimal one for the fundamental $SU(N)$ theories in the Veneziano limit. 
However, it turns out that the optimal truncation order estimated from the coefficients of the two largest orders 
was highly dependent on the details of the theories considered in this work. 
We therefore estimate the truncation error $\delta_1$ only in a practical sense, 
as defined in Eqs.~\ref{eq:delta_1} and \ref{eq:truncation_error}, 
but to be conservative enough to account for potentially severe nonperturbative effects 
implied by the ambiguity associated with the singularity in the Borel plane. 
In the former case, on the other hand, we first approximate the closed forms of $\girpsi$ and $\Girpsi$ 
using the Pad\'e approximation. 
The best approximant has been determined by comparing its qualitative behavior to the large $N_f$ expansion 
in a certain range of $N_f$ for which both calculations are supposed to be under control. 
The two definitions of the critical condition 
approximated by the best Pad\'e approximants 
generally lead to different values of $N_f^{\rm cr}$ 
and we take the difference as the uncertainty of our analysis which was denoted by $\delta_2$. 

In the large $N$ limit we basically found two different results for the conformal window, up to an overall factor of two, 
thanks to the large $N$ universality: one for the fundamental representation in the Veneziano limit, 
and the other for the adjoint and two-index representations in the 't Hooft limit. 
Our results show that for the former the two distinct errors, $\delta_1$ and $\delta_2$, 
are comparable to each other, 
but for the latter $\delta_2$ is about three times larger than $\delta_1$. 
At finite, but not too small $N$, we find that $\delta_2$ is much smaller than $\delta_1$ 
except for the fundamental $SU(N)$ and $Sp(2N)$ theories. 
For certain cases with small integer values of $N$ 
the truncation error $\delta_1$ turned out to be much larger than $\delta_2$ 
and prevented us from narrowing down the location of the lower edge of the conformal window. 
In these small $N$ theories, more dedicated studies with improved error analysis techniques 
and a better understanding of the critical condition 
will be helpful to further discriminate their IR nature near the phase boundary. 

Among various analytical methods found in the literature, 
we have considered three widely used ones to be compared with our results. 
For all the theories considered in \Sec{cw}, 
we find that our values of $N_f^{\rm cr}$ are typically larger than 
those estimated by the $2$-loop beta function, but smaller than those 
by the conventional Schwinger-Dyson analysis. 
Within the uncertainties estimated in this work, our results are in good agreement with those
determined by the all-order beta function analysis with $\girpsi=1$. Note that both approaches 
adopt the same critical condition defined in \Eq{critical_condition}. Nevertheless, such an agreement is 
somewhat surprising since the ways to determine $N^{\rm cr}$ are totally different: 
the all-order beta function analysis finds the IR fixed point using a relatively simple closed form of $\beta(\alpha)$, 
while our approach explicitly computes the anomalous dimension $\girpsi$ using the conformal expansion. 
Although nothing is yet conclusive, this result might in turn indicate that both the conjectured all-order beta function 
and the conformal expansion for $\girpsi$ well describe the conformal theory at the IR fixed point within the conformal window. 
Compared with recent lattice results, 
we find that our results of $N_f^{\rm cr}$ are in excellent agreement 
for $SU(3)$ with fermions in the fundamental and symmetric representations, 
and for $SU(2)$ with fermions in the fundamental representation. 
In the case of the adjoint $SU(2)$ our estimation for $N_f^{\rm cr}$ is somewhat larger than 
what has been found in lattice studies. 
Although further confirmation is required, such a discrepancy may reflect 
the limitation of our approach due to significant finite $N$ correction to the critical condition 
and/or 
sizable nonperturbative corrections to the conformal expansion at 
near the lower end of the conformal window in the small-$N$ theories. 
For $Sp(4)$ theories coupled to fundamental and antisymmetric fermions, 
which have received considerable attention in recent years, 
we predict that $N_f^{\rm cr}=9\sim 11$ and $5\sim6$, respectively. 
Besides the lattice methods, it would be also interesting to compare our results to bootstrap techniques, 
as the recent calculation of $\girpsi$ for the $N_f=12$ fundamental $SU(3)$ theory 
showed a promising result comparable to various lattice and analytical methods \cite{Li:2020bnb}. 

In this work we have continued our journey to the end of conformal window 
in nonsupersymmetric and asymptotically free gauge theories. 
To reach there 
we assume that the loss of conformality in the onset of chiral phase transition 
is featured by the critical condition to $\gpsi$ in \Eq{critical_condition}. 
We also assume that $\girpsi$ can perturbatively be calculated by the scheme-independent conformal expansion 
over the whole range of the conformal window. 
The determination of $N_f^{\rm cr}$ based on these assumptions, however, is challenged by 
our limited understanding on the nature of chiral phase transition 
and the potentially sizable nonperturbative effects. 
To address these issues, we have tried to estimate the errors in $N_f^{\rm cr}$ 
in two folds, as discussed in details in \Sec{error_estimate}, 
and argue that part of such systematic effects are well captured by our error estimates. 
In several finite $N$ theories, we also find that our values of $N_f^{\rm cr}$ 
are in good agreement with the recent lattice results. 
In this respect, 
we believe that our approach to the determination of the conformal window, along with the error analyses, 
would provide a useful analytical supplement and a guidance to more rigorous and better controlled nonperturbative calculations 
of (near) conformal theories. 

\acknowledgments
The author thanks M.~Serone and L.~Di Pietro for useful discussions about the critical condition 
to the anomalous dimension of a fermion bilinear and the large $N_f$ expansion. 
The author also thanks M.~Piai, S.~Rychkov and H.~Gies for comments and suggestions on the manuscript. 
This work is supported in part by the National Research Foundation of Korea grant funded by the Korea government(MSIT) (No.~NRF-2018R1C1B3001379) 
and in part by Korea Research Fellowship program funded by the Ministry of Science, ICT and Future Planning
through the National Research Foundation of Korea (No.~2016H1D3A1909283).

\bibliography{cwss}
\end{document}